\documentclass[fleqn,usenatbib]{mnras}

\usepackage{graphicx}
\usepackage{epstopdf,color,verbatim}
\usepackage{threeparttable}
\usepackage[]{txfonts}
\usepackage{amsbsy}
\def\simless{\mathbin{\lower 3pt\hbox
{$\rlap{\raise 5pt\hbox{$\char'074$}}\mathchar"7218$}}}   
\def\simmore{\mathbin{\lower 3pt\hbox
{$\rlap{\raise 5pt\hbox{$\char'076$}}\mathchar"7218$}}}   
\newcommand{\eqb}{\begin{eqnarray}}
\newcommand{\eqe}{\end{eqnarray}}
\newcommand{\be}{\begin{eqnarray}}
\newcommand{\ee}{\end{eqnarray}}
\newcommand{\bi}{\begin{itemize}}
\newcommand{\ei}{\end{itemize}}
\def\comp{\,c/\omega_{\rm p}}

\newcommand{\gmax}{\gamma_{n=4}}
\newcommand{\gsig}{\gamma_{\sigma}}
\newcommand{\gc}{\gamma_{\rm cut}}
\newcommand{\rL}{r_{\rm L}}

\newcommand{\sect}[1]{Sect.~\ref{#1}}

\newcommand{\fig}[1]{Fig.~\ref{fig:#1}}
\newcommand{\tab}[1]{Table~\ref{tab:#1}}
\newcommand{\eq}[1]{Eq.~(\ref{eq:#1})}

 \newcommand{\rhot}{\,r_{\rm L}}
\newcommand{\alf}{Alfv\'en}

\newcommand{\eB}{\epsilon_{\rm B}}
 
\title[The steady growth of the high-energy cutoff]
{The Steady Growth of the High-Energy Spectral Cutoff in Relativistic Magnetic Reconnection}

\author[Petropoulou \& Sironi]
{Maria Petropoulou$^1$\thanks{E-mail: m.petropoulou@astro.princeton.edu} 
 and Lorenzo Sironi$^{2}$\thanks{E-mail: lsironi@astro.columbia.edu}\\
$^1$Department of Astrophysical Sciences, Princeton University, 4 Ivy Lane, Princeton, NJ 08544, USA\\
$^2$Department of Astronomy, Columbia University, 550 W 120th St, New York, NY 10027, USA}

\begin{document}
\date{Received / Accepted}
\pagerange{\pageref{firstpage}--\pageref{lastpage}} \pubyear{2018}

\maketitle

\label{firstpage}

\begin{abstract}
Magnetic reconnection is invoked as an efficient particle accelerator in a variety of astrophysical sources of non-thermal high-energy radiation. With  large-scale two-dimensional particle-in-cell simulations of relativistic reconnection (i.e., with magnetization $\sigma\gg1$) in pair plasmas, we study the long-term evolution of the power-law slope and high-energy cutoff of the spectrum of accelerated particles. We find that the high-energy spectral cutoff does not saturate at $\gamma_{\rm cut}\sim 4\sigma$, as claimed by earlier studies, but it steadily grows with time as long as the reconnection process stays active. At late times, the cutoff scales approximately as $\gamma_{\rm cut}\propto \sqrt{t}$, regardless of the flow magnetization and initial temperature. We show that the particles dominating the high-energy spectral cutoff reside in plasmoids, and in particular in a strongly magnetized ring around the plasmoid core. The growth of their energy is driven by the increase in the local field strength, coupled with the conservation of the first adiabatic invariant. We also find that the power-law slope of the spectrum ($p=-{\rm d}\log N/{\rm d}\log \gamma$) evolves with time. For $\sigma\gtrsim10$, the spectrum is hard at early times ($p\lesssim 2$), but it tends to asymptote to  $p\sim 2$; the steepening of the power-law slope allows the spectral cutoff to extend to higher and higher energies, without violating the fixed energy budget of the system. Our results demonstrate that relativistic reconnection is a viable candidate for accelerating the high-energy particles emitting in relativistic astrophysical sources. 
\end{abstract} 

\section{Introduction}
A fundamental question in the physics of astrophysical relativistic outflows is 
how their energy, which is initially carried in the form of Poynting flux, is transferred to the plasma, and then radiated away to power the observed emission. Field dissipation via magnetic reconnection has been often invoked to explain the non-thermal signatures of pulsar wind nebulae \citep[PWNe; e.g.,][see \citealt{sironi_17} for a recent review]{lyubarsky_kirk_01,lyubarsky_03,kirk_sk_03,petri_lyubarsky_07,sironi_spitkovsky_11b,cerutti_12b, cerutti_13b,philippov_14,cerutti_17}, jets from active galactic nuclei \citep[AGNs; e.g.,][]{romanova_92,giannios_09,giannios_10b,giannios_13,petropoulou_16,christie_18} and gamma-ray bursts \citep[GRBs; e.g.,][]{thompson_94, thompson_06,usov_94,spruit_01,drenkhahn_02a,lyutikov_03,giannios_08}.  

In most relativistic astrophysical outflows, reconnection proceeds in the ``relativistic'' regime in which the magnetic energy per particle can exceed the rest mass energy (or equivalently, the magnetization $\sigma$ is larger than unity). The acceleration process of the radiating particles can only be captured from first principles by means of fully-kinetic particle-in-cell (PIC) simulations. Energisation of particles in relativistic reconnection of pair plasmas has been investigated in a number of PIC studies, both in two dimensions \citep[2D; e.g.,][]{zenitani_01,zenitani_07,jaroschek_04,bessho_07,bessho_12,hesse_zenitani_07,lyubarsky_liverts_08,cerutti_12b,ss_14,guo_14,guo_15a,liu_15,nalewajko_15,sironi_15,sironi_16,werner_16,kagan_16,kagan_18} and three dimensions \citep[3D; e.g.,][]{zenitani_05b,zenitani_08,liu_11,sironi_spitkovsky_11b, sironi_spitkovsky_12,kagan_13,cerutti_13b,ss_14,guo_15a,werner_17}.  Recently, 2D PIC simulations have started to tackle the acceleration capabilities of relativistic reconnection in electron-ion plasmas \citep[e.g.,][]{melzani_14,sironi_15,guo_16,rowan_17,werner_18,ball_18}.

In order to assess the role of relativistic reconnection as the process responsible for the non-thermal high-energy emission in astrophysical sources, it is important to quantify the properties of the energy spectrum of accelerated particles. This is typically modelled as a power law {$dN/d\gamma\propto \gamma^{-p}$ starting from a Lorentz factor $\gamma_{\min}$ and terminated by a high-energy cutoff at $\gc$ (typically,  $\gc \gg \gamma_{\min}$).} The power-law slope $p=-{\rm d}\log N/{\rm d}\log \gamma$ of the differential particle distribution has been shown to depend on the flow magnetization \citep{ss_14,guo_14,werner_16}, with harder spectra obtained for higher magnetizations. {For $\sigma\gtrsim 10$, where the power-law slope in pair plasmas is $p\lesssim 2$, the energy of the particle population ($\propto \int^{\gc}_{\gamma_{\min}} \!\! {\rm d}\gamma \, \gamma \, {\rm d}N/{\rm d}\gamma \propto \gc^{-p+2}$) is carried by the most energetic particles of the power law}. As the mean energy of particles accelerated by reconnection has to equal the average energy per particle in the pre-reconnection plasma (including the dominant magnetic contribution for $\sigma\gg1$), the finite energy budget of the system implies that the high-energy cutoff of the particle spectrum cannot extend to arbitrarily high energies, if $p\lesssim 2$.

Such energy-based constraint does not apply for $\sigma\lesssim 10$, where the power-law slope is steeper ($p\gtrsim 2$), and low-energy particles dominate both the number ($\propto \gamma_{\min}^{-p+1}$) and the energy census ($\propto \gamma_{\min}^{-p+2}$). However, recent studies of relativistic reconnection in pair plasmas \citep[][see also \citealt{guo_16}, for electron-ion plasmas]{werner_16, kagan_18} claimed that, even for the steep spectra found at $\sigma\lesssim 10$, the high-energy cutoff of the particle distribution does not extend beyond a Lorentz factor of $\sim 4 \sigma$. More precisely, \citet{werner_16} found that the high-energy cutoff is super-exponential for small computational boxes, where it scales linearly with the system size, and it becomes exponential for sufficiently large boxes, where it equals $\sim 4\sigma$, regardless of box size. {Originally, the limit of $4\sigma$ was claimed for reconnection in electron-positron plasmas; in an electron-proton plasma, where the magnetization is defined with respect to the proton rest-mass energy density, similar arguments would lead to a (claimed) upper limit on the proton Lorentz factor of $\sim 4\sigma$ and on the electron Lorentz factor of  $\sim 4 (m_{\rm i}/m_{\rm e})\sigma$.} At face value, {these results would pose} a major problem for models of reconnection-powered particle acceleration, since virtually all astrophysical sources require non-thermal particles with energies well beyond this limit.

The goal of this work is to revisit this claim, by means of 2D PIC simulations of relativistic reconnection in pair plasmas. Our unprecedentedly large computational domains allow us to study the long-term evolution of the system, well beyond the initial transient phase. We find that: (\textit{i}) the high-energy cutoff of the particle spectrum does not saturate at $\gamma_{\rm cut}\sim 4\sigma$, but it steadily grows to significantly larger values, as long as the reconnection process stays active; the evolution is fast at early times (up to $\gamma_{\rm cut}\sim 4\sigma$) and slower at later times, which might have been incorrectly interpreted as a saturation of the cutoff at $\gamma_{\rm cut}\sim 4\sigma$; (\textit{ii}) at late times, the cutoff scales approximately as $\gamma_{\rm cut}\propto \sqrt{t}$, regardless of the flow magnetization and the temperature of the pre-reconnection plasma; (\textit{iii}) this scaling is measured not only for the whole reconnection region, but also for plasmoids that remain ``isolated'', i.e., when they do not experience mergers with other plasmoids of comparable sizes; this suggests that mergers are not the main drivers of the evolution of the high-energy cutoff; (\textit{iv}) at any given time, the particles controlling the high-energy spectral cutoff reside in plasmoids, and in particular in a strongly magnetized ring around the plasmoid core; (\textit{v}) by following the trajectories of a large number of high-energy particles, we find that the growth of their energy (and so, of the spectral cutoff) is driven by the increase in magnetic field at the particle location 
coupled with the conservation of the first adiabatic invariant;  mergers provide the ground for multiple energization/compression cycles, by scattering the high-energy particles from the island cores towards the outskirts, where the field is weaker;
(\textit{vi}) we also find that the power-law slope softens over time: for $\sigma=10$, it asymptotes to $p\sim 2$, corresponding to equal energy content per logarithmic interval in Lorentz factor; for $\sigma=50$, the power law index grows with time from $p\sim1.2$ up to $p\sim 1.7$, and we argue that at even later times it may asymptote to the same value as for $\sigma=10$, i.e., $p\sim 2$; the steepening of the power-law slope over time observed for $\sigma=50$ allows the spectral cutoff to extend to higher and higher energies, without violating the fixed energy budget of the system (see the argument presented above).

This paper is organised as follows. In Sect.~\ref{sec:setup} we describe the setup of our simulations. In Sect.~\ref{sec:structure} we present the temporal evolution of the structure of the reconnection layer. In Sect.~\ref{sec:results} we focus on the evolution of the particle energy spectrum, illustrating how the power-law slope and high-energy spectral cutoff depend on time. The physical origin of the increase in the high-energy spectral cutoff is explained in Sect.~\ref{sec:origin}. In Sect.~\ref{sec:discussion} we summarise our results and present the main astrophysical implications of our findings. We conclude in Sect.~\ref{sec:concl}.

\section{Simulation setup}\label{sec:setup}
We use the 3D electromagnetic PIC code TRISTAN-MP \citep{buneman_93, spitkovsky_05} to study magnetic reconnection in pair plasmas.  We explore anti-parallel reconnection, i.e., we set the guide field perpendicular to the alternating fields to be zero. The reconnection layer is initialised as a Harris sheet with the magnetic field $\bmath{B}=-B_0\,\tanh\,(2\pi y/\Delta)\,\bmath{\hat{x}}$ reversing at $y=0$ over a thickness $\Delta$. The thickness of the current sheet is chosen to be small enough  (see \tab{param}) as to make the sheet tearing unstable, even in the absence of an initially imposed perturbation. So, in contrast to most studies, we investigate untriggered (i.e., spontaneous) reconnection, where the tearing mode starts from numerical noise. 

The field strength $B_0$ is defined through the magnetization $\sigma=B_0^2/4\pi h$, where $h$ is the enthalpy density of the unreconnected plasma  including the contribution of its rest-mass energy density, i.e. $h=n_0 m c^2 + \hat{\gamma} n_0 k_B T_0/(\hat{\gamma}-1)$; here $c$ is the speed of light, $m$ is the electron mass, $n_0$ is the number density {(including both particle species)}, $T_0$ is the temperature of the unreconnected plasma, and {$\hat{\gamma}=4/3$ is the adiabatic index of a relativistically hot plasma. For a cold plasma, the enthalpy density is dominated by the rest mass, and the second term is unimportant.} The \alf\ speed is related to the magnetization as $v_A/c=\sqrt{\sigma/(\sigma+1)}$. We focus on the regime of relativistic reconnection (i.e., $v_A/c\sim 1$) and explore cases with $\sigma=10$ and 50 (see \tab{param}). The plasma outside the layer is taken to be cold with a small thermal spread upon initialisation ($k_B T_0/m c^2 \equiv \Theta_{\rm e}=10^{-4}$), but we also explore a few cases with hot upstream plasma having $\Theta_{\rm e}=3$ (see \tab{param}). The magnetic pressure outside the current sheet is balanced by the particle pressure in the sheet. This is achieved by adding a component of hot plasma with over-density $\eta=3$ relative to the number density $n_0$ of particles outside the layer. 
We exclude the hot particles initialised in the current sheet from the particle energy spectra we show in the following sections, as their properties depend on our choice of the sheet initialisation. 

We perform simulations on a 2D spatial domain, but we track all three components of the velocity and of the electromagnetic fields. For a direct comparison to previous studies, we adopt periodic boundary conditions in the $x$ direction of the reconnection outflow \citep[e.g.,][]{ss_14, werner_16, kagan_18}. We also employ two moving injectors receding from $y=0$ at the speed of light along $\pm {\bmath{\hat{y}}}$ (i.e., the direction of the reconnection inflow) as well as an expanding simulation box. The two injectors constantly introduce fresh magnetized plasma into the simulation domain. This permits us to evolve the system for long enough times, while keeping all the regions that are in causal contact with the initial setup. This choice has clear advantages over the double-periodic setup that is commonly employed \citep[e.g.,][]{werner_16}, where the limited amount of particles and magnetic energy will necessarily inhibit the evolution of the system to long times. This technique  has been extensively employed in studies of relativistic shocks \citep{sironi_spitkovsky_09,sironi_spitkovsky_11a,sironi_13} and more recently extended to studies of relativistic reconnection \citep{ss_14, sironi_16, rowan_17, ball_18}.

To investigate the long-term evolution of particle acceleration in reconnection we performed simulations in unprecedentedly large-scale computational domains. Henceforth, we adopt as our  unit of length the Larmor radius $\rL$ of particles with energy $\sim \sigma m c^2$, implicitly assuming that the magnetic energy is fully transferred to particles. Then, $\rL$ is related to the plasma skin depth $\comp=\sqrt{m c^2/4\pi n_0 e^2}$, as $\rhot=\sqrt{\sigma}\comp$. For a relativistically hot upstream plasma, the skin depth definition includes an additional factor of $\sqrt{1+3\Theta_{\rm e}}$, to account for the effects of relativistic inertia. We choose a spatial resolution of $\comp$=5 or 10 cells which corresponds to a temporal resolution of several time steps per one Larmor gyration period $2\pi/\sqrt{\sigma}\,\omega_{\rm p}$, even for the largest magnetization $\sigma=50$ that we explored\footnote{The numerical speed of light is 0.45 cells per time step.}. We typically employ 4 particles per cell, but we also performed a simulation with 16 particles per cell and found no differences (see \tab{param}).  The size of the computational domain along the reconnection layer $L$ ranges from hundreds to thousands of $\rhot$, as presented in \tab{param}. {The box size along the $y$ direction increases over time and by the end of the simulation it is comparable or larger than the $x$ extent.} As we show in the following sections, such large domains are of paramount importance to attain a sufficient dynamic range for studying the evolution of the high-energy cutoff of the  particle energy spectrum, without being artificially constrained by the boundary conditions. 

\begin{table}
 \centering 
 \caption{Physical and numerical parameters of the simulations presented in this study.}
  \begin{threeparttable}
 \begin{tabular}{c c c c c c c}
  \hline 
 Run &  $\comp$ & $\sigma$ & $\Theta_{\rm e}$ & $\Delta/\rL$ & $L/\rL $ & Duration $[\rL/c]$\tnote{\textdagger}\\
 \hline 
 R[a]&	5	& 10	&   $10^{-4}$      &  3           & 1062	& 2158	\\
 R[b]\tnote{*}&5	& 10& $10^{-4}$	   &  6           & 1062	& 2158	\\
 R[c]&	5	& 10	&   $10^{-4}$	   &  9           & 1062	& 2158/3557	\\
 R[d]&	5	& 10	&   $10^{-4}$	   &  6           & 531		& 1091	\\
 R[e]&	5	& 10	&   $10^{-4}$	   &  6           & 2125	& 4292/5806\\
 R[f]&	5	& 10	&   $10^{-4}$	   &  6           & 4250	& 8561	\\
 R[g]&	5	& 10	&   3		   &  6           & 531		& 1091	\\
 R[h]&	5	& 10	&   3		   &  6           & 1062	& 2158	\\
 R[i]&	5	& 10	&   3		   &  6           & 2125	& 4292	\\
 R[j]&	5	& 50	&   $10^{-4}$	   &  6		  & 950		& 4292	\\
 R[i]&	10	& 10	&   $10^{-4}$	   &  6		  & 531		& 1079	\\
 R[k]&	10	& 10	&   $10^{-4}$	   &  6		  & 1062	& 2146	\\
 \hline 
 \label{tab:param}
 \end{tabular}
 \begin{tablenotes}
 \item[\textdagger]The duration in light crossing times, $[L/c]$, can be obtained by dividing the last column by the second-to-last one.
 \item[*]Four additional simulations were performed with either a low-order integration scheme for Maxwell's equations (``loword''), or a higher number of particles per cell (``ppc16''), or different degrees of filtering of the currents (``ntimes8'' and ``ntimes32'') -- see \fig{num} in Appendix~\ref{sec:app1}.  
\end{tablenotes}
 \end{threeparttable}
\end{table}
\section{The structure of the reconnection layer}\label{sec:structure}
\begin{figure*}
\centering
 \resizebox{\hsize}{!}{\includegraphics{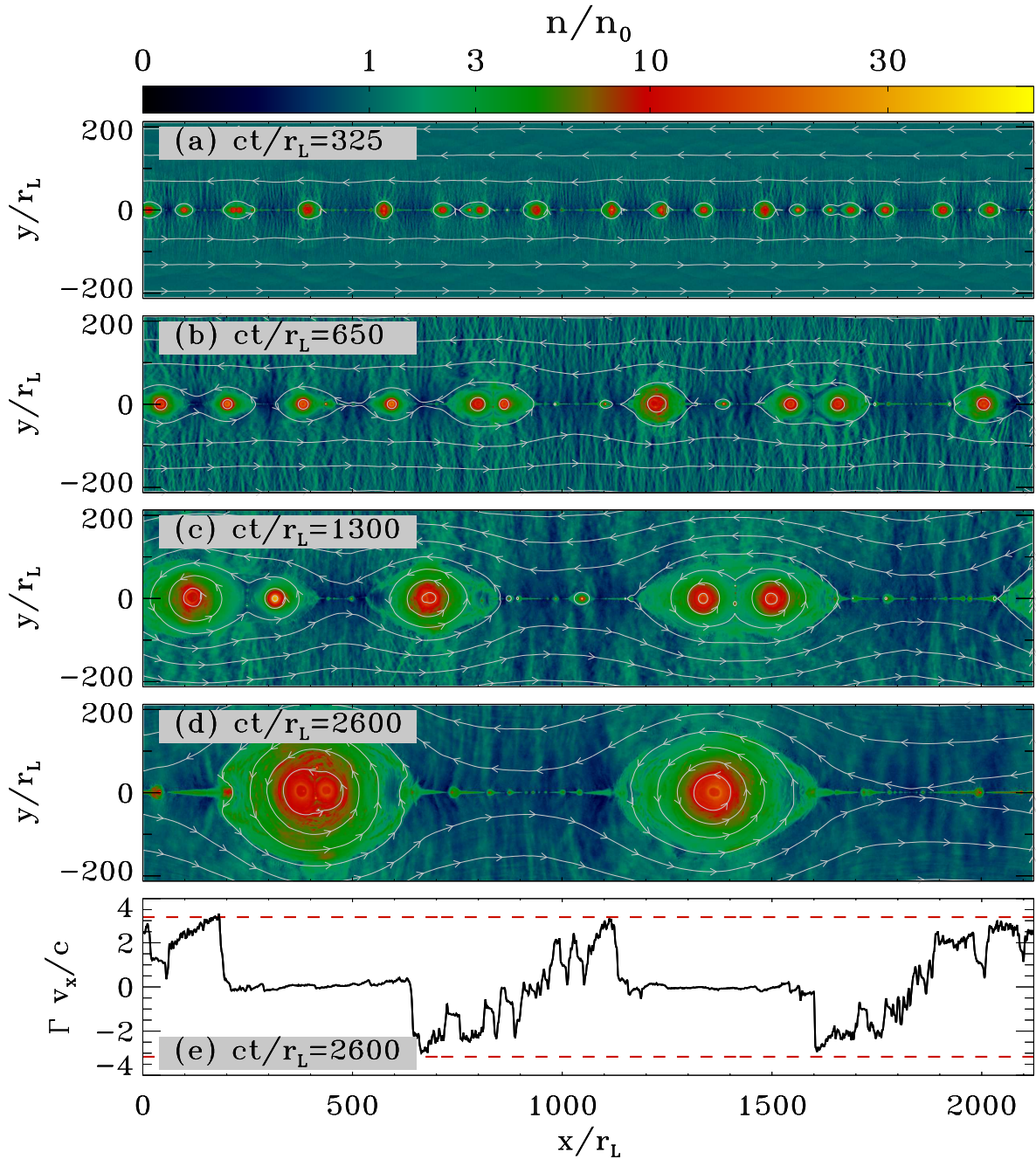}}
 \caption{2D structure of the particle number density $n$, in units of the number density $n_0$ far from the reconnection layer, from a simulation with $\sigma=10$ (R[e] in \tab{param}). We only show the region $|y|/L<0.15$ to emphasise the small-scale structures in the reconnection layer (the extent of the computational box along $y$ increases at the speed of light, as described in \sect{sec:setup}). The 2D density structure at different times (as marked on the plots) is shown in the panels from top to  bottom, with overplotted magnetic field lines. The spatial dependence of the plasma outflow 4-velocity (i.e., along the $x$ direction at $y=0$) computed at $ct/r_{\rm L}=2600$ is also shown in the bottom panel. The horizontal dashed lines denote the asymptotic values $\pm \sqrt{\sigma}$ predicted by analytic theory of relativistic magnetic reconnection \citep{lyubarsky_05}. }
 \label{fig:fluidtime}
\end{figure*}
Upon the onset of reconnection, the layer becomes unstable to the primary tearing mode and breaks into a series of magnetic islands (primary plasmoids) separated by X-points and secondary current sheets \citep[e.g.,][]{furth_63, loureiro_07, uzdensky_16, comisso_16}. The secondary current sheets, in turn, fragment into a chain of smaller islands (secondary plasmoids), as a result of the secondary tearing mode or ``plasmoid instability'' discussed by \cite{uzdensky_10}. Both primary and secondary islands move along the layer, coalesce with other neighbouring ones, and grow to larger scales. Despite the similarities in their evolution, primary and secondary plasmoids have different internal structures. The former have cores with negligible magnetic content, which are supported by the pressure of the hot particles initially present in the current sheet \citep[see e.g.][]{sironi_15, nalewajko_15}; the field strength peaks in a ring surrounding the hot core. In contrast to primary islands, secondary plasmoids do not bear memory of the current sheet initialisation (for their structure, see Appendix A in \citealt{sironi_16}). Since the highest energy particles that are of interest for this study reside in primary plasmoids, below we focus mostly on primary plasmoids and discuss their particle content and structure in the following sections (see \sect{sec:results} and \sect{sec:origin}). {Henceforth, we will refer to primary plasmoids simply as {\it plasmoids}, unless stated otherwise.}

The temporal evolution of the reconnection region is illustrated in \fig{fluidtime}, where we show snapshots of the 2D structure of the particle number density from one of our large-scale simulations in a $\sigma=10$ pair plasma (R[e] in \tab{param}). At late times (see e.g. panel (d)), the reconnection region is dominated by a few large primary plasmoids which will eventually merge forming a bigger one whose size is comparable to the layer's length. The formation of such a large plasmoid in combination with the periodic boundary conditions will eventually inhibit the inflow of fresh plasma in the layer, thus shutting off the reconnection process. We have verified that the reconnection process remains active for the entire duration of all simulations listed in \tab{param}. This is of paramount importance for addressing the question of whether the apparent saturation in the high-energy cutoff of the particle spectrum \citep[see][]{werner_16, kagan_18} is a mere consequence of the drop in reconnection rate induced by the periodic boundaries, or whether it holds even when reconnection is allowed to keep operating in steady state (see \sect{sec:results}).

The bottom panel of \fig{fluidtime} displays the spatial dependence of the plasma's bulk outflow 4-velocity (i.e., along the direction of the layer)  at the same time as panel (d). The 4-velocity drops to zero at the locations occupied by large plasmoids, as these are in good approximation stationary structures. In the regions in between primary plasmoids, however, the 4-velocity of outflowing plasma approaches the asymptotic value $\sqrt{\sigma}$ predicted by  analytic theory of relativistic magnetic reconnection \citep{lyubarsky_05} and numerically demonstrated in PIC simulations of relativistic reconnection with outflow boundary conditions \citep{sironi_16}. In other words, the region in between two primary plasmoids has the same dynamics as in outflow simulations. This suggests that the plasma outflow in relativistic reconnection can achieve its asymptotic velocity  independently from the initialisation of the layer, from the way of initiating reconnection (untriggered versus triggered setup), and from the boundary conditions.  This would not have been possible though, if the computational domain employed in the periodic untriggered runs presented in this work was not sufficiently large. 
 
\begin{figure}
\centering
 \resizebox{\hsize}{!}{\includegraphics[trim=0 0 0 0, clip=true]{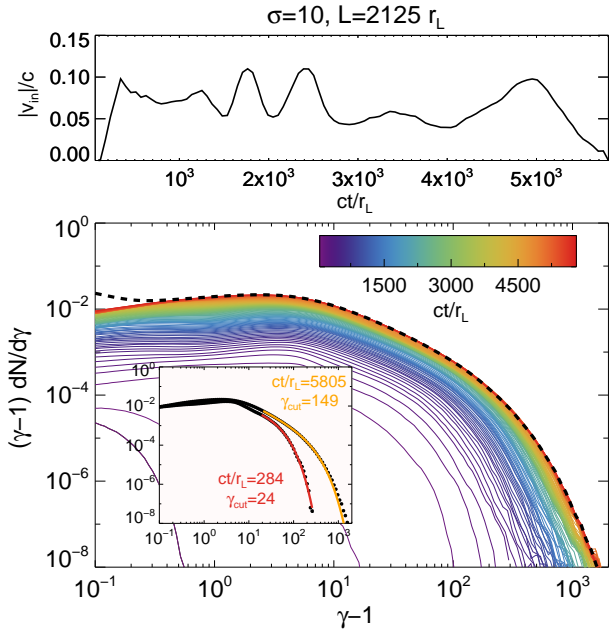}}
 \caption{Top panel: plot of the inflow plasma speed (in units of the speed of light) as a function of time (in units of $\rL/c$) for a $\sigma=10$ simulation of reconnection (R[e] in \tab{param}). The layer's length is $L=2125\, \rL$. The inflow speed is averaged over a slab of width $0.2\, L$ across the layer (i.e., along the $y$ direction), but our results are nearly insensitive to this choice. Bottom panel: temporal evolution of the particle spectrum $(\gamma-1) {\rm d}N/{\rm d}\gamma$ extracted from the reconnection downstream region -- see also \fig{fluidtime}, for a depiction of the layer structure. The particle spectrum from the whole box at the end of the simulation is also shown (black dashed line). Particle spectra are normalised to the total number of particles at the end of the simulation. The inset shows two indicative snapshots of the particle energy spectrum (black symbols) with the fitting results overlaid with coloured lines. The values of the cutoff Lorentz factor, $\gc$, at the corresponding times are also marked on the inset plot. A movie showing the evolution of the layer structure and particle distribution can be found at \url{https://youtu.be/0SwViHBo_s4}.}
 \label{fig:spec}
\end{figure}

\begin{figure}
\centering
 \resizebox{\hsize}{!}{\includegraphics[trim=0 20 0 0, clip=true]{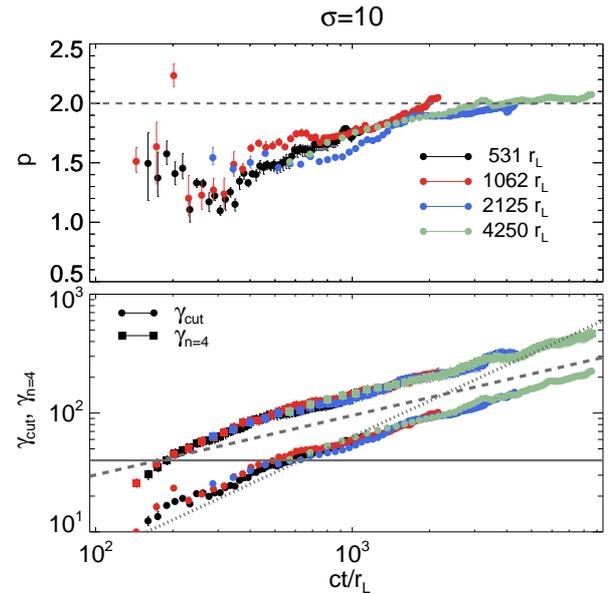}}
 \caption{Top panel: temporal evolution of the power-law slope $p$ of the particle distribution (${\rm d}N/{\rm d}\gamma\propto \gamma^{-p}$). The horizontal dashed line corresponds to $p=2$, yielding equal energy content per decade in Lorentz factor. Bottom panel: temporal evolution of the cutoff Lorentz factor $\gc$ (circles) and the maximum Lorentz factor $\gmax$ defined in eq.~(\ref{eq:gmax}) (squares) of the particle distribution. Two scalings of the Lorentz factors with time are overplotted for comparison, namely $\propto t$ (dotted line) and $\propto \sqrt{t}$ (dashed line). {The horizontal solid line marks the value of $4\sigma$.} Coloured symbols denote different sizes of the simulation domain (in units of $\rL$), as marked on the upper panel (see R[b], R[d]-R[f] in Table~\ref{tab:param}). {In all cases, the reconnection remains active for the displayed time interval.}
 }
 \label{fig:boxsize}
\end{figure}

\section{The steady growth of the high-energy spectral cutoff}\label{sec:results}
We are interested in studying the temporal evolution of the energy spectrum of particles from the region where the plasma has undergone reconnection.  To identify this ``reconnection downstream region'' (hereafter, simply ``reconnection region'') we used a mixing criterion, as proposed by \citet{daughton_14}. Particles are tagged with an identifier based on their initial location with respect to the current sheet (i.e., $y$ at $t=0$ above or below the sheet). Particles from these two regions are mixed in the course of the reconnection process. We can thus identify as the reconnection region the ensemble of computational cells with mixing fraction above a certain threshold \cite[for more details, we refer the reader to][]{rowan_17, ball_18}. 

As an indicative example, we show the temporal evolution of the particle energy spectrum from the reconnection region in \fig{spec} (bottom panel). The structure of the reconnection layer at different times is shown in \fig{fluidtime}. The spectrum evolves quickly at early times (dark blue lines), resulting in a broad non-thermal tail already by $t\sim 300 \, \rL/c$, as also found by  \citet{ss_14, werner_16, kagan_18}. Although the spectral evolution appears to be more gradual at later times, i.e.  $t\gtrsim 10^3 \, \rL/c$,  the highest energy part of the spectrum keeps extending to higher and higher values of the particle Lorentz factor. During this period of steady growth, the reconnection process remains active, as illustrated in \fig{spec} (top panel), where we show the temporal evolution of the reconnection rate (averaged over the layer's length). The reconnection rate ranges between $0.05\,c$ and $0.10\,c$ and it begins to decline only at $t\gtrsim 5\times 10^3 \, \rL/c$, or equivalently at $\gtrsim 2.3\, L/c$ (when the reconnection process starts to shut off, due to the periodic boundary conditions in the $x$ direction). The speed of plasma flowing into the layer fluctuates over time due to the motion and coalescence of large primary plasmoids. For example, peaks in the reconnection rate are associated with the virulent growth of the secondary tearing mode in the long layers that stretch in between two receding primary plasmoids. On the other hand,  
the reconnection rate drops after a merger, since the merger produces outward propagating waves that tend to decelerate the upstream flow. 
 
After the strong spectral evolution at early times, the particle energy spectrum can be  described by a power law of index $p$ and normalisation $N_0$ with a high-energy exponential cutoff at $\gc$:
\be
f(\gamma)\equiv\frac{dN}{d\gamma} = N_0 \left(\frac{\gamma}{\gamma_*}\right)^{-p} e^{-\frac{\gamma}{\gc}}, \quad \gamma \ge \gamma_*,
\label{eq:dndg}
\ee 
where $\gamma_*=20-30$ in all cases studied here.  
This is illustrated in the inset  plot of \fig{spec}, where we show fits (coloured lines) to the particle energy spectrum (black symbols) for two different times marked on the plot. The high-energy cutoff increases from $\sim 20$ at $ct/r_{\rm L}\sim 300$ up to $\sim 150$ at $ct/r_{\rm L}\sim 5800$. The results presented in \fig{spec} are in tension with the findings of past studies \citep{werner_16, kagan_18}, where the high-energy cutoff of the particle spectrum was found to saturate at $\sim 4 \sigma\sim 40$.
We argue that this was an artificial result of the limited extent of the computational domains employed in these studies, as we demonstrate below.

We performed a suite of 2D simulations of reconnection in pair plasma with $\sigma=10$ and different sizes of the computational domain ranging from $L=531 \, \rL$ to $L=4250 \, \rL$ (see R[b], R[d]-R[f] in \tab{param}). These values should be compared to the sizes of the largest simulations performed by \cite{werner_16} and \cite{kagan_18} which are, respectively, $200\, \rL$ and $320 \, \rL$, for values of $\sigma$ similar to what we adopt here. \fig{boxsize} presents the temporal evolution of the power-law index $p$ (top panel) and the cutoff $\gc$ (bottom panel) obtained by fitting the particle energy spectra of the reconnection  region at different times with expression (\ref{eq:dndg}). The high-energy cutoff is also compared against  {a model-independent proxy of the maximum Lorentz factor of the particle spectrum} \citep[see also][]{bai_15}:
\be 
\gamma_{n} = \frac{\int \, {\rm d}\gamma \, \gamma^{n+1} f(\gamma)}{\int \, {\rm d}\gamma \, \gamma^n f(\gamma)}.
\label{eq:gmax}
\ee 
{Here, $n=4$ is chosen so that $\gmax$ can probe the highest energy part of the particle spectrum, while still not being limited by small number statistics.} The main results shown in \fig{boxsize} are:
\bi
\item the high-energy cutoff increases steadily with time as long as the reconnection process remains active, which is true for all the cases shown in \fig{boxsize} (with a box of length $L$ resulting in active reconnection for a few $L/c$). 
\item the fast early-time evolution of $\gc$ is followed by a slower temporal increase starting at $t\sim 600 \, \rL/c$. At the time of the transition, we find that $\gc \sim 4\sigma$. At later times, the high-energy cutoff clearly exceeds the $4\sigma$ limit discussed by \citet{werner_16} and \citet{kagan_18}, even for the smallest simulation we performed ($L=531 \, \rL$). The late-time growth of $\gc$ and $\gmax$ scales, in good approximation, as $\propto \sqrt{t}$.  Thus, it is possible that  in previous studies the transition to a slower phase of growth was interpreted as cessation/saturation of the spectral evolution.
\item $\gc$ and $\gmax$ show the same time dependence, although the former is obtained from a fit (i.e., it is model-dependent) and the latter is not (see eq.~(\ref{eq:gmax})). This suggests that the adopted fitting function in eq.~(\ref{eq:dndg}) is a good choice for describing the data. 
\item at all times, we find that $\gmax \approx 3 \gc$. This is in agreement with the relation $\gmax \approx (n-p+1) \gc$, which is obtained from eq.~(\ref{eq:gmax}) upon substitution of  eq.~(\ref{eq:dndg}). For our adopted $n=4$ and the slope $p\sim 2$ measured from the fit (see top panel), the relation $\gmax \approx 3 \gc$ is then naturally expected.
\item the power-law segment of the particle energy spectrum becomes progressively softer over time, as the power-law index increases from $p\sim 1.5$ at early times to $p\sim 2$ at late times (top panel). The early-time large variations of $p$ are a result of the limited extent of the power-law segment that we use for the fit (see also the inset plot in \fig{spec}). The late-time value of $p$, which seems to asymptote to $p\sim 2$, is consistent with that quoted by \citet{ss_14} for $\sigma=10$ (see Fig.~3 therein). Harder spectra than those reported here and in \cite{ss_14} had been quoted by \cite{werner_16} and \cite{guo_14}, for the same magnetization. This is also likely to be a consequence of a limited box size and/or simulation time span of those studies.
\ei 
\subsection{Dependence on the physical conditions}
In this section, we explore the dependence of the results presented in \fig{boxsize} on the physical conditions of the layer, namely the initial thickness of the current sheet $\Delta$, the initial temperature of the unreconnected plasma $\Theta_{\rm e}$, and the plasma magnetization $\sigma$ (as defined with the enthalpy density). 

\begin{figure}
\centering
 \resizebox{\hsize}{!}{\includegraphics[trim=0 20 0 80, clip=true]{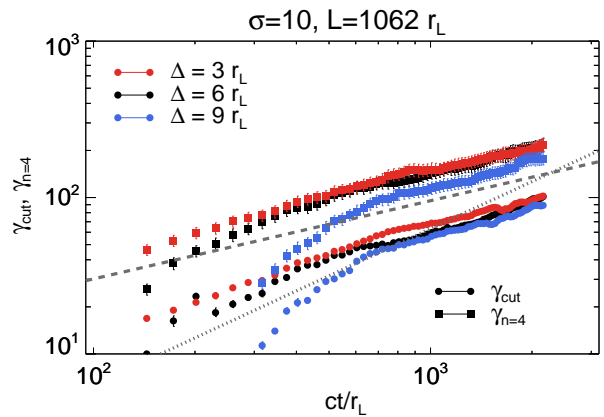}}
 \caption{Plot of the cutoff (circles) and maximum (squares) Lorentz factors of the particle distribution as a function of time for three different choices of the layer's thickness $\Delta$ marked on the plot (see R[a]-R[c] in Table~\ref{tab:param}). The dashed and dotted lines have the same meaning as in \fig{boxsize}. }
 \label{fig:thickness}
\end{figure}

We performed simulations where we vary the layer's thickness, as shown in \tab{param} (R[a]-R[c]), while keeping the layer's length fixed at $L=1062\, \rL$. The temporal evolution of $\gc$ (circles) obtained by fitting the particle energy spectra for different current sheet thicknesses is shown in \fig{thickness}. For comparison, we also show $\gmax$ (squares) computed according to eq.~(\ref{eq:gmax}). The thickness of the current sheet affects only the early-time evolution of $\gc$ and $\gmax$ (i.e., $t\lesssim 600 \, \rL/c$). The onset of the tearing instability is expected to be delayed for thicker current sheets \citep{zenitani_07}. As a result, both $\gc$ and $\gmax$ reach higher values for smaller $\Delta$ within the same time interval. For $t \gtrsim 600 \, \rL/c$ though, the temporal evolutions of the high-energy cutoff and of the maximum Lorentz factor are independent of $\Delta$. In fact, the late-time increase of $\gc$ and $\gmax$ follows a $\propto \sqrt{t}$ scaling, as also shown in \fig{boxsize}. 

\begin{figure}
 \centering 
 \resizebox{\hsize}{!}{\includegraphics[trim=0 20 0 80, clip=true]{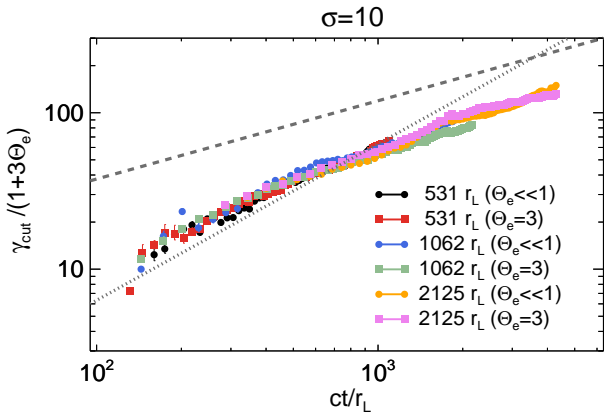}}
 \caption{Temporal evolution of the cutoff Lorentz factor of the particle distribution in the case of cold ($\Theta_{\rm e} \ll 1$) and hot ($\Theta_{\rm e}=3$) pair plasmas with $\sigma=10$ (see R[b], R[d]-R[e], and R[g]-R[i] in Table~\ref{tab:param}). Here, $\gc$ is normalised to $1+3\Theta_{\rm e}$, i.e., to the mean Lorentz factor of upstream particles. Results obtained for different sizes of the simulation domain are plotted with different colours (see inset legend). The dashed and dotted lines have the same meaning as in \fig{boxsize}. } 
 \label{fig:delgam}
\end{figure}
\begin{figure}
\centering
 \resizebox{\hsize}{!}{\includegraphics[trim=10 30 0 0, clip=true]{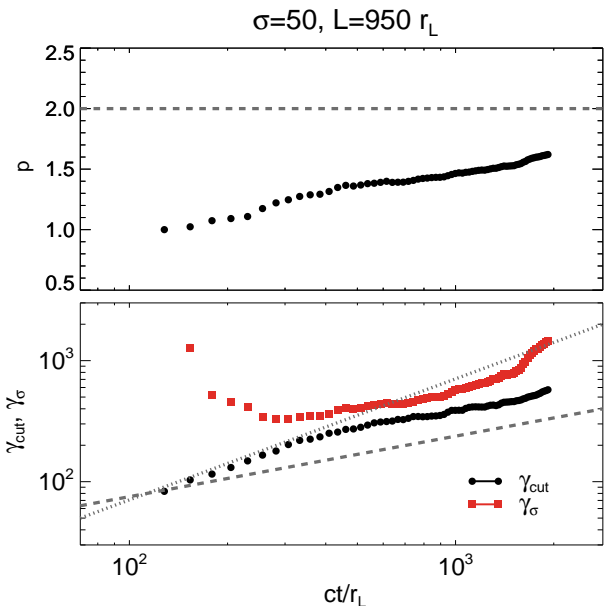}}
 \caption{Top panel: plot of the particle distribution's power-law slope $p$ as a function of time, for reconnection in pair plasma with $\sigma=50$ (R[j] in \tab{param}). The horizontal dashed line corresponds to $p=2$, yielding equal energy content per decade in Lorentz factor.  Bottom panel: plot of the cutoff Lorentz factor of the particle distribution $\gc$ (black circles) as a function of time. The maximum Lorentz factor allowed by the finite energy budget of the system $\gsig$ (see eq.~(\ref{eq:gsig})) is overplotted for comparison (red squares). The dashed and dotted lines have the same meaning as in \fig{boxsize}.
 }
 \label{fig:sig50}
\end{figure}
We also carried out simulations of reconnection in a magnetized ($\sigma=10$) hot upstream pair plasma with initial temperature $\Theta_{\rm e}=3$ and three different sizes of the layer's length (R[g]-R[i]). Our results are compared to those from cold plasma simulations of the same size (respectively, R[b], R[d]-R[e]) in \fig{delgam}, where we show the temporal evolution of the high-energy cutoff $\gc$. When this is normalised to 
$1+3\Theta_{\rm e}$, i.e., to the mean Lorentz factor\footnote{{The mean Lorentz factor of upstream plasma is defined as $1+ \Theta_{\rm e}/(\hat{\gamma}-1)$. Here, $\hat{\gamma}=4/3$ is the appropriate adiabatic index of a relativistically hot plasma; for a cold plasma, the mean Lorentz factor is close to unity.}} of upstream particles, all curves overlap.  Although not shown here, we find that the entire high-energy non-thermal spectra overlap. This confirms that the physics of relativistic pair reconnection does not depend on $\Theta_{\rm e}$ as long as the magnetization and plasma skin depth are defined in a way to include the effects of relativistic inertia (see \sect{sec:setup}), see also \citet{lyutikov_17}.  

Finally, we studied the long-term evolution of the particle energy spectrum from a simulation of reconnection in cold plasma with $\sigma=50$ (see R[j] in ~\tab{param}). The power-law index and the high-energy cutoff obtained from fitting the data are presented in \fig{sig50} (top and bottom panels, respectively). The particle energy spectrum is hard (i.e., $p<2$), in agreement with previous studies \citep{ss_14, guo_14,guo_15a,werner_16}.  A spectral slope $p<2$ would not allow the spectrum to extend to arbitrarily high energies. The fact that the mean energy per particle is $\sim (1 + \sigma/2)\, mc^2$ implies that a power-law energy spectrum with slope $p$ starting from $\gamma\sim 1$ would extend at most up to a limiting Lorentz factor $\gsig$ \citep[see also][]{ss_14, werner_16}: 
\be
\gsig \approx \left[\left(1+\frac{\sigma}{2}\right)\frac{2-p}{p-1} \right]^{1/(2-p)}.
\label{eq:gsig}
\ee 
Interestingly, we find a slow but steady evolution of the power-law index towards larger values (i.e., steeper spectra) as the simulation evolves.  At this point, we cannot exclude the possibility that the slope will saturate at $p\sim 2$, i.e., the same value to which it asymptotes for $\sigma=10$ (see \fig{boxsize}). This would imply that relativistic reconnection in plasmas with $\sigma >10$ would eventually produce particle spectra with  equal energy per logarithmic energy band, despite the fact that the early-time spectra can be hard ($p<2$).

The progressive softening of the power law leads to larger values of $\gsig$ which, in turn, implies that the high-energy cutoff of the particle energy spectrum is also allowed to increase (still, with the constraint $\gamma_{\rm cut}< \gamma_{\sigma}$). This is exemplified in the bottom panel of  \fig{sig50}, where we show the temporal evolution of $\gc$ (black circles) and $\gsig$ (red squares). The latter is computed according to eq.~(\ref{eq:gsig}) for $p$ values derived from the fit (top panel in \fig{sig50}). The high-energy cutoff Lorentz factor $\gamma_{\rm cut}$ is at all times smaller than the limiting Lorentz factor $\gsig$. The rapid increase of $\gsig$ seen at $t\sim 1800 \,\rL/c$ is a result of $p \rightarrow 2$. \fig{sig50} clearly shows that there is no energy crisis for $\sigma \gg 1$, i.e., the evolution of the spectrum to arbitrarily large Lorentz factors  is not inhibited by the finite energy budget of the system, thanks to the simultaneous steepening of the power law.

\section{The origin of the steady growth}\label{sec:origin}
By employing a suite of large-scale 2D simulations of reconnection in pair plasmas for different numerical and physical parameters, we have confidently demonstrated  that the high-energy cutoff of the particle energy spectrum exceeds the $4\sigma$ limit and that it increases steadily as $\gc \propto \sqrt{t}$ (see \sect{sec:results}). We now proceed to explain the origin of this steady growth by looking in more detail to the properties of individual plasmoids and particles. For this purpose, we will be using simulation R[b] in \tab{param} throughout this section. {We note that the structure of the layer and plasma dynamics (e.g., outflow velocity) of simulation R[b] are similar to those discussed in Sect.~\ref{sec:structure} and shown in Fig.~\ref{fig:fluidtime}.}

\begin{figure}
\centering 
 \resizebox{\hsize}{!}{\includegraphics[trim=0 0 0 0, clip=true]{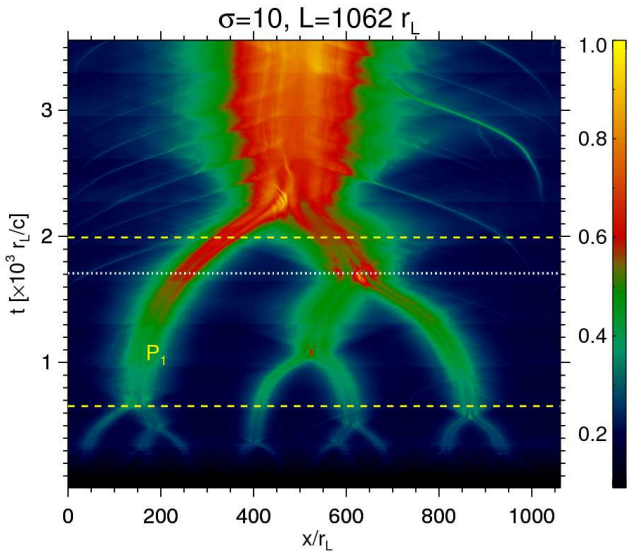}} 
\caption{Time-position diagram showing the magnetic energy fraction normalised to its maximum value (colour bar) from a simulation of reconnection in pair plasma with $\sigma=10$ (R[b] in \tab{param}). The plasmoid $P_1$ forming at $x\sim 150\, \rL$ and $t\sim 700 \, \rL/c$ remains ``isolated'' for most of the simulation duration, as it does not merge with other primary plasmoids until $t\sim 2\times 10^3\, \rL/c$. The time interval when it remains isolated is delimited by the two dashed yellow lines. The dotted white line indicates the time used for showing the structure of the plasmoid $P_1$ and of the layer in Figs.~\ref{fig:tomoisl} and \ref{fig:hrmap}, respectively.
 }
 \label{fig:posttime}
\end{figure}
\subsection{Plasmoid tomography}\label{sec:island}
\fig{posttime} shows a time-position diagram of the reconnection layer coloured according to the magnetic energy fraction $\epsilon_{\rm B}$ defined as $\eB=B^2/8\pi n_0 mc^2$, where $B$ is the local magnetic field strength. The magnetic energy fraction is normalised to its maximum value obtained throughout the simulation domain and duration. The temporal evolution of the layer can be adequately described as a merger tree whose branches denote the life tracks of plasmoids. At early times (e.g., $t\lesssim 500\, \rL/c$), many small-sized primary plasmoids are born along the layer due to the tearing instability. They later merge with each other to form bigger islands, like the one 
represented by the leftmost branch in \fig{posttime} for times between $\sim 600\,\rL/c$ and $\sim 2\times10^3\, \rL/c$. During this time interval (which is delimited by the horizontal dashed lines in \fig{posttime}), this plasmoid, which we hence call $P_1$, can be considered ``isolated'', as it does not undergo major mergers (i.e., mergers with other primary plasmoids). Still, our definition does not exclude mergers with secondary plasmoids that are constantly created in the layer. At late times (after $2.2\times 10^3\,\rL/c$), we find that the layer is dominated by a single plasmoid with size $\gtrsim 0.3 L$, which attracts all the newly created secondary islands (see the thin green-coloured branches in \fig{posttime}, for $t\gtrsim 2.2\times 10^3\,\rL/c$). The magnetic energy density increases during the lifetime of each primary plasmoid, as indicated by the variation in colour (e.g., from green to red) along individual branches. 


\begin{figure}
 \centering 
  \resizebox{\hsize}{!}{\includegraphics[trim=0 20 0 0]{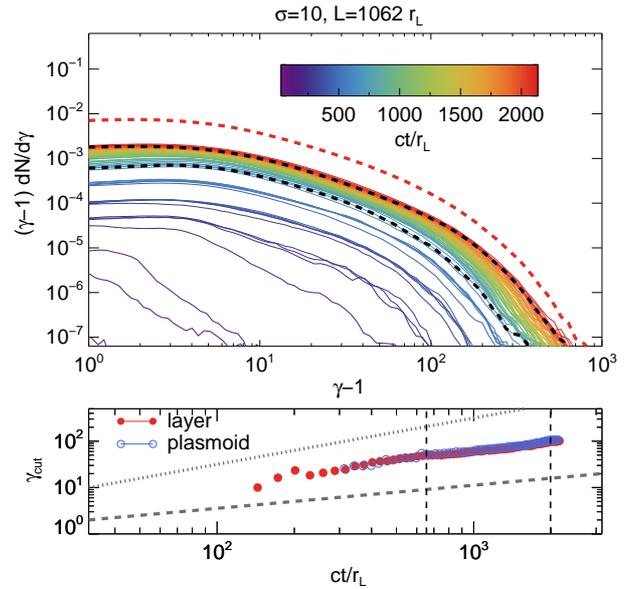}} 
 \caption{Top panel: temporal evolution of the particle spectrum $(\gamma-1)\,{\rm d}N/{\rm d}\gamma$  (coloured curves) of plasmoid $P_1$ identified in \fig{posttime}. The particle spectrum from the whole box at the end of the simulation is also shown (red dashed line). All spectra are normalised to the total number of particles at the end of the simulation. Spectra computed during the period when the plasmoid $P_1$ does not undergo major mergers (i.e., it is isolated) are delimited by the two black dashed lines (at times corresponding to the two dashed yellow lines in \fig{posttime}).  Bottom panel: plot of the cutoff Lorentz factor $\gc$ as a function of time (in units of $\rL/c$), as obtained by fitting the particle energy spectrum from the whole reconnection  region (filled symbols) and from the plasmoid $P_1$ (open symbols). This plasmoid is isolated for the time window denoted with dashed vertical lines. Grey-coloured lines have the same meaning as in \fig{boxsize}.
 } 
 \label{fig:islspec}
\end{figure}

\begin{figure}
 \centering 
 \resizebox{\hsize}{!}{\includegraphics{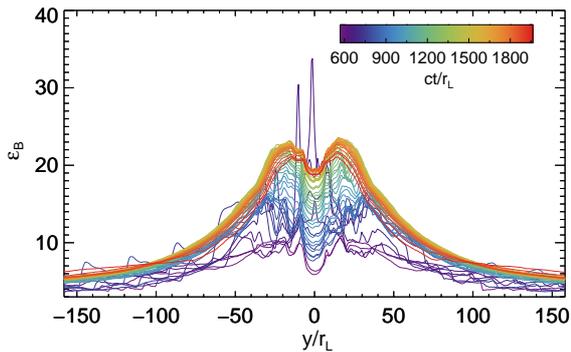}}
 \caption{Transverse profiles of the magnetic energy fraction $\epsilon_{\rm B}$ for the plasmoid $P_1$ identified in \fig{posttime}. Different colours correspond to times when the plasmoid is isolated (see inset colour bar) -- see also the two horizontal dashed yellow  lines  in \fig{posttime}, delimiting the time interval when the plasmoid is isolated. The magnetic energy density close to the plasmoid centre increases with time, while it remains approximately constant at the plasmoid periphery, at the upstream value $\sim \sigma/2$.} 
 \label{fig:eBnomajor}
\end{figure}

Overall, \fig{posttime} shows that the reconnection region at $t\gtrsim 10^3\, \rL/c$ is dominated by a few primary plasmoids, including $P_1$. They contain a significant fraction of the total number of particles residing in the reconnection region. It is therefore interesting to compare the particle energy spectrum from the whole reconnection  region with that of an individual primary plasmoid, {which is identified using contours of the vector potential \citep[for details, see Sect.~2.1 of][]{sironi_16}. \fig{islspec} shows} the temporal evolution of the particle energy spectrum of plasmoid $P_1$ (top panel), from its birth until the time when it merges with the only other primary plasmoid left in the layer ( at $t\sim 2.2\times 10^3\,\rL/c$). At early times, plasmoid $P_1$ undergoes several mergers with other plasmoids of similar sizes. Here, its spectral evolution is discontinuous, as evidenced by the jumps between consecutive curves (see the blue-coloured lines). On the contrary, the spectral evolution in between major merger episodes (and in particular, during the isolated phase at $650 \lesssim ct/\rL\lesssim 2\times 10^3$) is smooth and gradual (see the coloured lines enclosed by the two dashed black lines). Importantly, the particle spectrum extends to higher and higher energies even during the phase when the plasmoid stays isolated, i.e., mergers are not a necessary condition for the spectral evolution. This is also illustrated in the bottom panel of \fig{islspec}, where we show that during the isolated phase (delimited by the two vertical dashed lines), the high-energy cutoff in the  spectrum of plasmoid $P_1$ steadily grows, still as $\gamma_{\rm cut}\propto \sqrt{t}$ (compare with the oblique dashed grey line).
 
The particle energy spectrum of the whole reconnection  region (red dashed line, referring to $t\sim 2.2\times 10^3\,\rL/c$) appears to differ only by a normalisation factor from that of the plasmoid $P_1$ (red solid line, referring to the same time). Indeed, the high-energy cutoff obtained by fitting the plasmoid spectrum is almost identical to that inferred from the spectrum of the entire reconnection  region (bottom panel in \fig{islspec}). Given the similarity of the particle energy spectra of the entire reconnection region and of plasmoid $P_1$, we can focus on the latter when trying to pinpoint the origin of the steady growth of the high-energy cutoff. 

\begin{figure*}
 \centering 
  \includegraphics[width=0.47\textwidth]{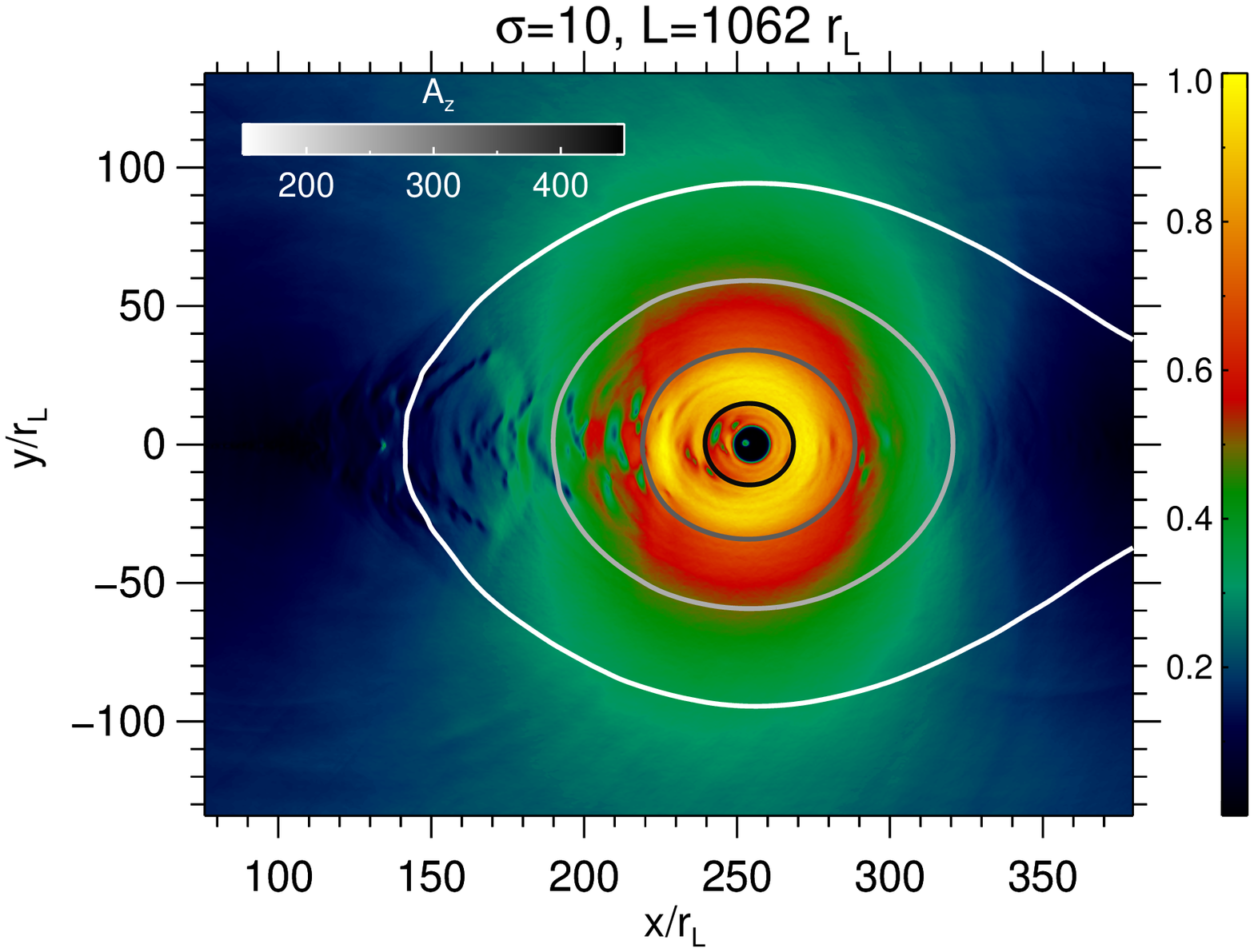}
  \hfill
  \includegraphics[width=0.47\textwidth]{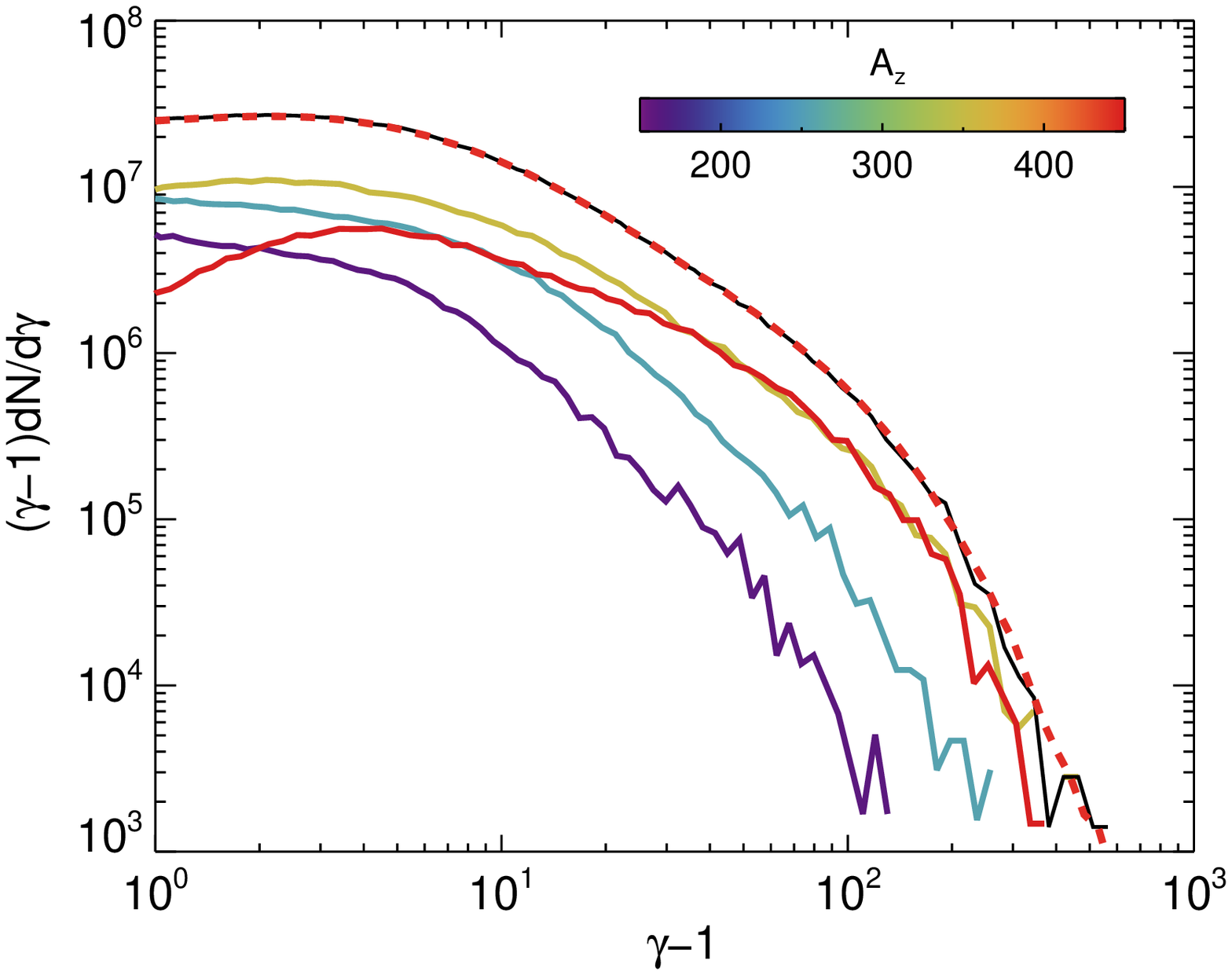}
 \caption{Left panel: 2D structure of the magnetic energy density normalised to its maximum value (vertical colour bar) in the vicinity of the isolated plasmoid $P_1$ at $t=1708 \, \rL/c$ with overlaid contours of the vector potential $A_z$ (see inset colour bar). Right panel: spatial dependence of the isolated plasmoid particle distribution $(\gamma-1){\rm d}N/{\rm d}\gamma$ (coloured lines) at $t=1708 \, \rL/c$.  Each curve shows the particle spectrum from an annulus delimited by two consecutive contours of magnetic vector potential (see the colorbar, and compare with the left panel). The particle spectrum from the whole plasmoid (red dashed line) matches perfectly the sum of the spectra from the annuli (black solid line).} 
 \label{fig:tomoisl}
\end{figure*}

\begin{figure*}
 \centering 
 \resizebox{\hsize}{!}{\includegraphics{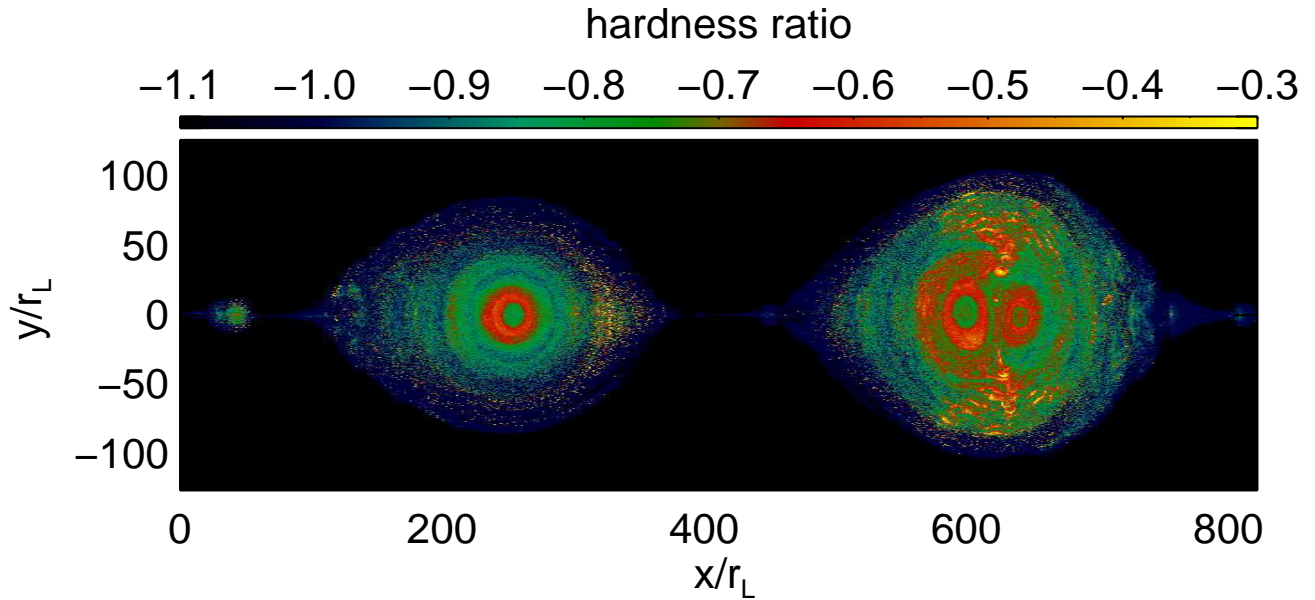}} 
 \caption{2D map of the hardness ratio, as defined by eq.~(\ref{eq:hr}), at $t=1708\, \rL/c$. More negative values of the hardness ratio correspond to softer particle energy spectra and {\it vice versa}. Regions devoid of particles with $\gamma \ge 5$ are artificially assigned to $HR<-1$ (black-coloured regions). }
 \label{fig:hrmap}
\end{figure*}

First, we consider the temporal evolution of the magnetic energy fraction in plasmoid $P_1$, see \fig{eBnomajor}. More specifically, we show how the transverse  profile (i.e., across the layer) of $\eB$ in plasmoid $P_1$ changes with time, during the interval when the plasmoid is not affected by major mergers. As time progresses, the profile of $\eB$ shows a double-peaked shape, which is suggestive of a ring with higher magnetic energy density around the plasmoid core {at a distance of $\sim 20 \, \rL$} (see  also Figs.~\ref{fig:tomoisl} and \ref{fig:hrmap}). The magnetic energy density in the ring increases over time as the plasmoid grows and accretes more magnetic flux, while it remains approximately constant at the plasmoid  periphery, at the upstream value $\sim\sigma/2$. Although not shown explicitly, the transverse density profile follows  a similar trend (see also \fig{fluidtime}).  

Let us take next a closer look at the spatial dependence of the particle energy spectrum within plasmoid $P_1$. To do so, at $t\sim 1708\, c/r_{\rm L}$ (denoted also in \fig{posttime} by a horizontal dotted line) we extract the particle energy spectrum from four concentric annuli defined by contours of the vector potential $A_z$, as shown in the left panel of \fig{tomoisl} (compare the inset legend in both panels). Colours in the left panel indicate the 2D structure of the magnetic energy density normalised to its maximum value in plasmoid $P_1$ (see vertical colour bar). On the one hand, the particle energy spectra from the two outer annuli are soft (i.e., have large $p$ value) and extend up to moderate Lorentz factors, $\gamma\sim 200$ (see right panel, purple and cyan lines). On the other hand, the spectra extracted from the two inner annuli (gold and red lines in the right panel) are harder and extend up to higher energies, $\gamma \sim  600$. In fact, particles residing close to the plasmoid centre (more specifically, particles residing in the magnetized ring) are those that determine the high-energy cutoff of the entire plasmoid spectrum, which is overplotted with a red dashed line for comparison. {These particles do not go through the inner non-magnetized core, but they stay confined to the magnetized ring, whose thickness increases as more particles are accreted.} In summary, we conclude that (\textit{i}) the spectral cutoff of the plasmoid spectrum is controlled by particles gyrating in the magnetized ring, and that (\textit{ii}) as shown in \fig{eBnomajor}, the field strength in the magnetized ring steadily grows over time.

\fig{tomoisl} shows that the spectral hardness of the particle energy spectrum depends on the location within the plasmoid. This is better illustrated in \fig{hrmap}, where we provide a late-time overview ($t=1708\, \rL/c$) of the whole reconnection layer in terms of the hardness ratio:
\be
HR=\frac{N_{\rm h}-N_{\rm s}}{N_{\rm h}+N_{\rm s}},
\label{eq:hr}
\ee
where $N_{\rm s}$ and $N_{\rm h}$ are the numbers of particles in the ``soft'' ($5 < \gamma \le 25$) and ``hard'' ($25 < \gamma \le 250$) energy bands, respectively. With this definition, the more negative the hardness ratio becomes, the softer the particle energy spectrum is and {\it vice versa}. Although the hardness ratio given by eq.~(\ref{eq:hr}) is bounded between -1 and 1, we artificially assign $HR$ values below -1 to regions devoid of particles with $\gamma>5$ (black-coloured regions in \fig{hrmap}).

The $HR$ map in \fig{hrmap} provides a detailed view of a plasmoid structure, as regard to the particle spectrum. Plasmoid $P_1$ is located at the left hand-side of the layer ($150 \, \rL \lesssim  x\lesssim 400 \, \rL$) and has not undergone major mergers since $t\sim 700\, \rL/c$. A distinctive feature is the presence of a ``hard'' ring (red-coloured region) surrounding the plasmoid core (green-coloured filled circle), in agreement with the results shown in \fig{tomoisl} (right panel). This also corresponds to the ring of high $\eB$ shown in \fig{eBnomajor} and \fig{tomoisl} (left panel). The dominant presence of particles with $\gamma$ up to 250 (i.e., above $\gc$) in the inner regions of an isolated plasmoid requires that a local mechanism for particle energisation be at work; as we show below, the particles are well magnetized, so they remain tied to the magnetic field lines, and they need to be accelerated {\it in situ}. The plasmoid outskirts are typically softer (green and blue colours), with some outer patches appearing harder (yellow colours) than the inner regions. These hard outer regions are most likely populated by particles accelerated during mergers with secondary plasmoids. Still, the high-energy cutoff of the entire plasmoid spectrum is not controlled by the energetic particles at the outskirts, since they are nearly irrelevant for the number census (see also right panel in \fig{tomoisl}). 

Similar features can be identified in the structure of the two plasmoids that are undergoing a major merger at the right hand-side of the layer. Notice also the presence of hard regions along the interface of the two merging plasmoids (red and yellow colours), suggesting that particles are also appreciably accelerated  during major mergers \citep[e.g.][]{nalewajko_15}.  However, major mergers are not necessarily important for driving the temporal evolution of the high-energy cutoff, given that --- as we have demonstrated above --- in isolated (i.e., merger-free) plasmoids the high-energy cutoff $\gamma_{\rm cut}$ is found to evolve in the same exact way as in the whole layer.

\begin{figure*}
 \centering 
  \resizebox{0.8\hsize}{!}{\includegraphics{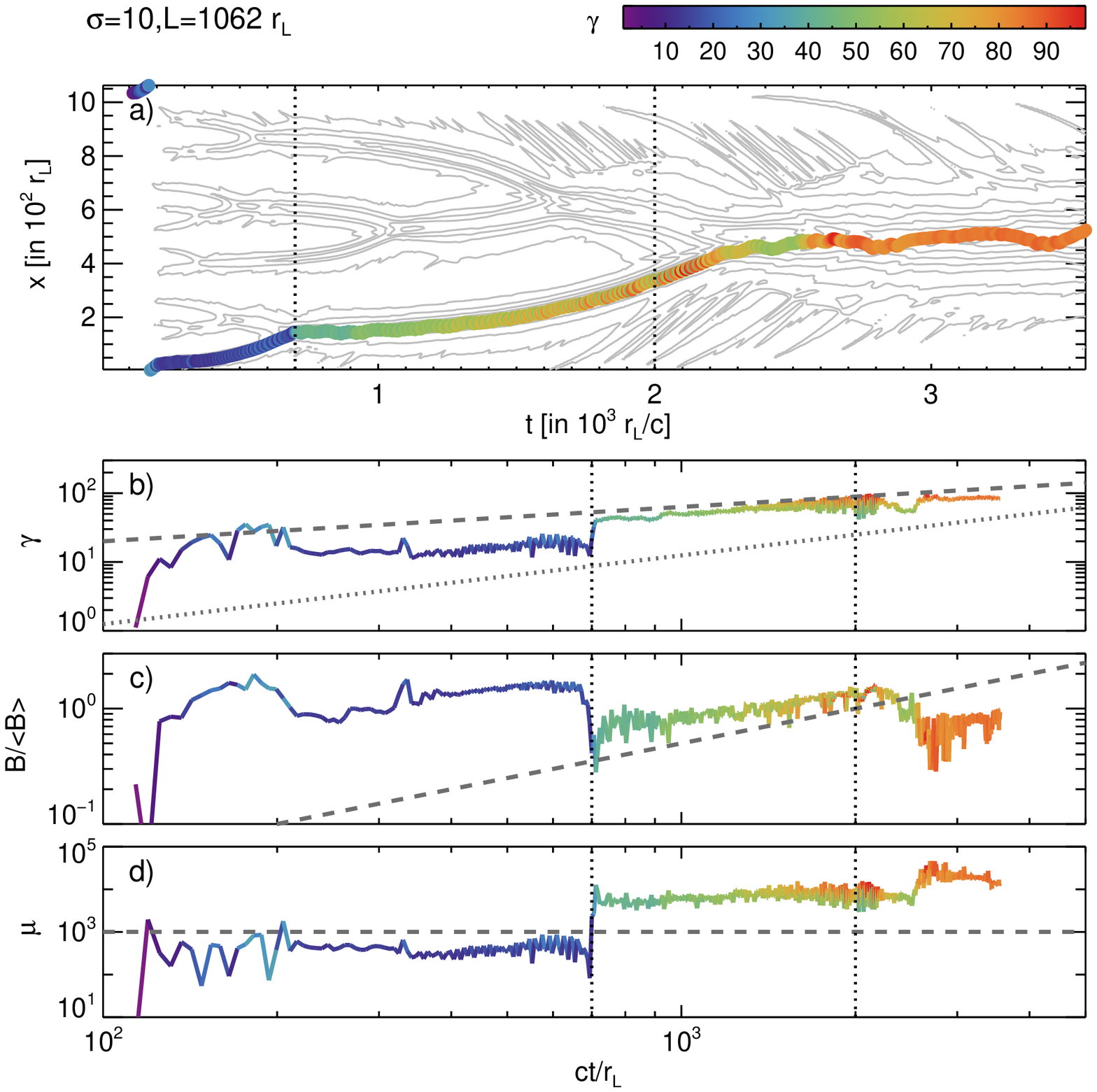}} 
 \caption{Plots showing the temporal evolution of various quantities for a representative positron that gets trapped in the isolated plasmoid $P_1$ shown in \fig{posttime}. From top to bottom we show: the position-time diagram of the particle, with contours of relativistic  ($\gamma \ge 1.05$) positron density (in arbitrary units) overlaid (panel (a)); the particle's Lorentz factor $\gamma$  as a function of time (panel (b)), with the dashed line scaling as $\propto \sqrt{t}$ and the dotted one as $\propto t$; the magnetic field strength $B$ at the particle's location (normalised to its lifetime average),  as a function of time (panel (c)), with the dashed line scaling as $\propto t$; the temporal evolution of the particle's magnetic moment $\mu$ (panel (d)). All curves are colour coded based on the particle Lorentz factor (see colour bar at the top). The positron is initially located on the right hand-side of the layer ($x\sim 10^3 \rL$), but it ``reappears'' on the other side at $t\sim 150 \, \rL/c$, due to the periodicity of the computational box in the $x$ direction. In panel (a), the time axis is in linear units, whereas we use logarithmic units in the other panels. In all panels, vertical dotted lines  indicate the times of two major mergers; in between, the plasmoid $P_1$ is isolated. A movie of the particle energy evolution and motion in the layer can be found at \url{https://youtu.be/e_5-mJE32Q0}.} 
 \label{fig:multi_prt140}
\end{figure*}

\subsection{Particle tracking}\label{sec:particles}
In \sect{sec:island} we showed that the evolution of the high-energy cutoff in  a large isolated plasmoid is similar to that from the entire reconnection  region. Furthermore, the highest-energy part of the plasmoid particle spectrum is dominated by particles residing in a ring of high $\eB$ around the plasmoid core. Here, we select an ensemble of particles that get trapped in plasmoid $P_1$ at some point of their lifetime, and reach high energies ($\gamma \geq 4\sigma$) by $t\sim2000 \,\rL/c$ (when the plasmoid ceases to be isolated). This allows to study the temporal evolution of their properties, such as their Lorentz factor, in a statistical way, and clarify the physical origin of their energisation.

This is exemplified in \fig{multi_prt140}, where we show the trajectory of a representative particle (positron) in the time-position diagram of the layer (panel (a)). Colours indicate the particle Lorentz factor (see colour bar at the top). The particle is initially located on the right hand-side of the layer ($x\sim 10^3 \rL$), but it ``reappears'' on the other side at $t\sim 150 \, \rL/c$, due to the periodicity of the computational box in the $x$ direction. After this point, the particle gets captured in a plasmoid that merges with a neighbouring one at $\sim 700\,\rL/c$ (leftmost vertical dotted line) to form plasmoid $P_1$ (see also \fig{posttime}). At the time of the merger, the particle's Lorentz factor increases suddenly from $\sim10$ to $\sim 40$, as shown in panel (b).  The large energy increase within a short time interval is consistent with particle acceleration by the anti-reconnection electric field at the interface between the two merging plasmoids \citep{ss_14, nalewajko_15}. At the time of the merger, the particle is scattered to the outer regions of the newly formed plasmoid, where the magnetic field is weaker (see e.g., \fig{eBnomajor}). This is also indicated by a sharp decrease of the magnetic field $B$ at the particle location (see panel (c)). Particle acceleration during mergers is a non-adiabatic process, where the particle adiabatic invariant is broken. In fact, during the merger the particle magnetic moment $\mu= m\gamma^2v_\perp^2/ 2 B${, where $v_{\perp}$ is the particle's velocity perpendicular to the local magnetic field $B$,} is not conserved  (panel (d) at $t\sim 700\,\rL/c$). Instead it increases at the time of the merger, because the particle momentum increases and the magnetic field at the particle location decreases.

Following this early-time merger, the plasmoid $P_1$ remains isolated for more than $10^3 \, \rL/c$ (the isolated phase is delimited by the two vertical dotted lines in \fig{multi_prt140}). During this time interval,  the particle is confined close to the plasmoid centre (panel (a)), while its Lorentz factor increases gradually, scaling approximately as $\propto \sqrt{t}$ (see dashed line in panel (b)). As we show below, such a scaling is common to most of the high-energy particles, which naturally explains the temporal evolution of the high-energy spectral cutoff.
As a result of accretion of magnetic flux and particles, the plasmoid compresses and the magnetic field strength at the particle location increases by a factor of $\sim 2$ from $t\sim 800 \, \rL/c$ to $t\sim 2\times10^3 \rL/c$, i.e., almost linearly with time (see panel (c) and \fig{eBnomajor}).  Meanwhile, $\mu$ is approximately constant. This implies that $\gamma v_\perp \approx \gamma c \propto \sqrt{t}$, as indeed shown in panel (b)\footnote{We have verified that most particles have perpendicular momentum appreciably larger than their parallel momentum.}.  At $t \sim 2\times 10^3 \, \rL/c$, the two large islands left in the layer (see also \fig{hrmap}) begin to merge. Because of their large sizes, the merger lasts longer as compared to earlier mergers. This is also indicated by the gradual change of $\gamma, B$, and $\mu$ starting at $\sim 2\times 10^3\, \rL/c$. The decrease in $\gamma$ between $2.2\times10^3 \, \rL/c$ and $2.5 \times 10^3 \, \rL/c$ results from the fact that, for this positron, the anti-reconnection electric field in between the two merging plasmoids is anti-aligned with respect to the particle velocity (i.e., $v_z\cdot E_z<0$). As it is demonstrated here, mergers do not always lead to an energy increase.

\begin{figure}
\centering
 \includegraphics[width=0.49\textwidth]{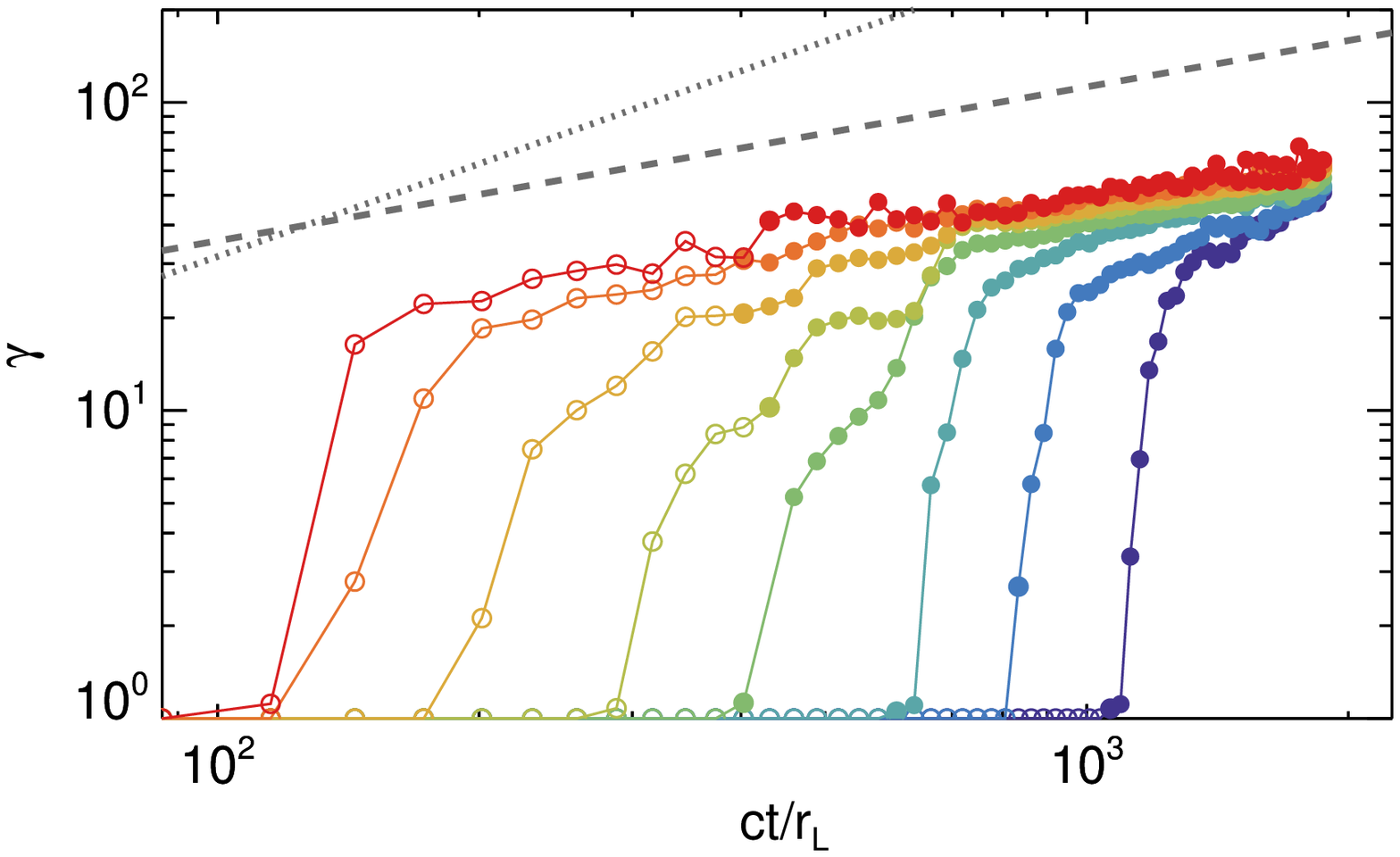}
 \includegraphics[width=0.49\textwidth]{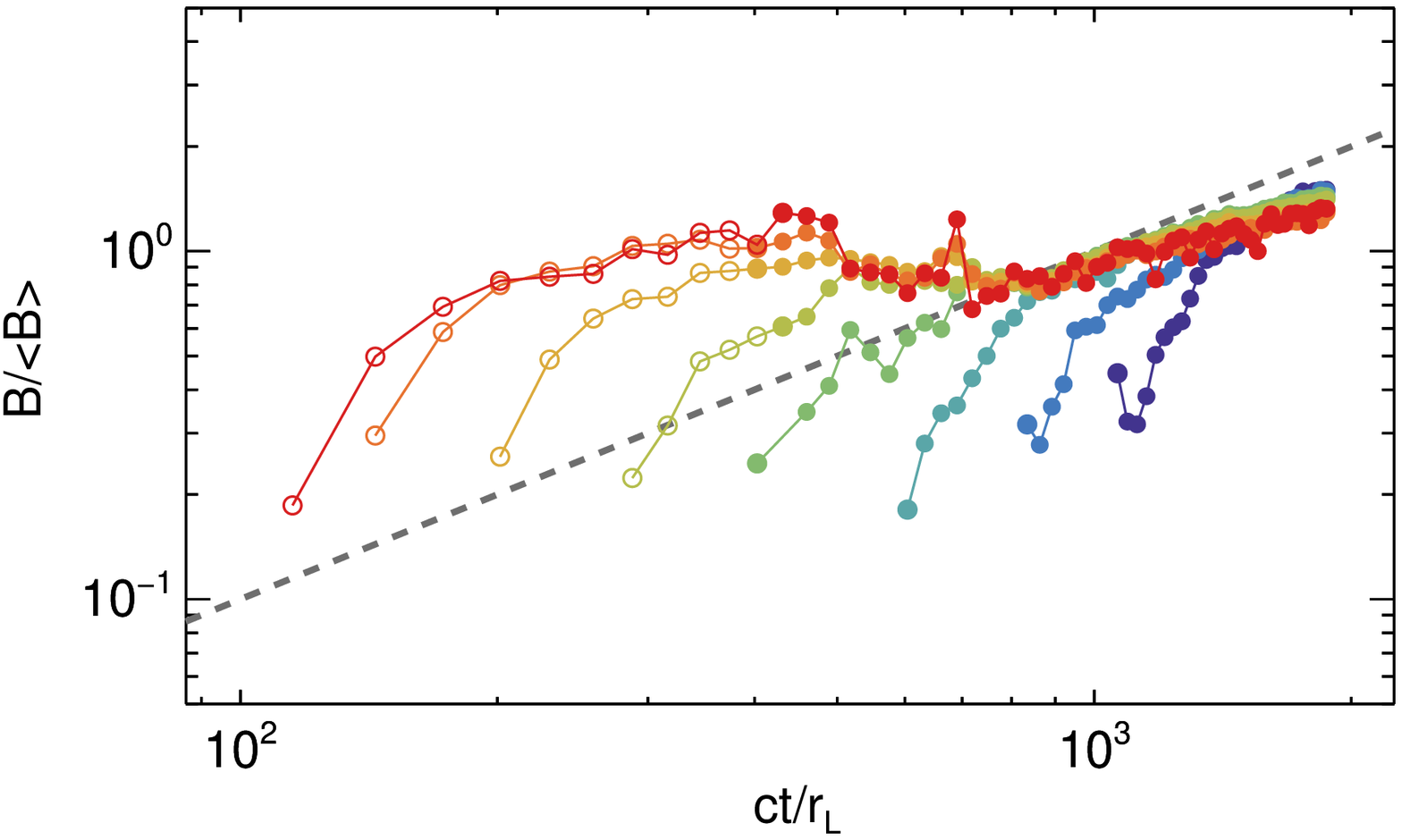}
 \includegraphics[width=0.49\textwidth]{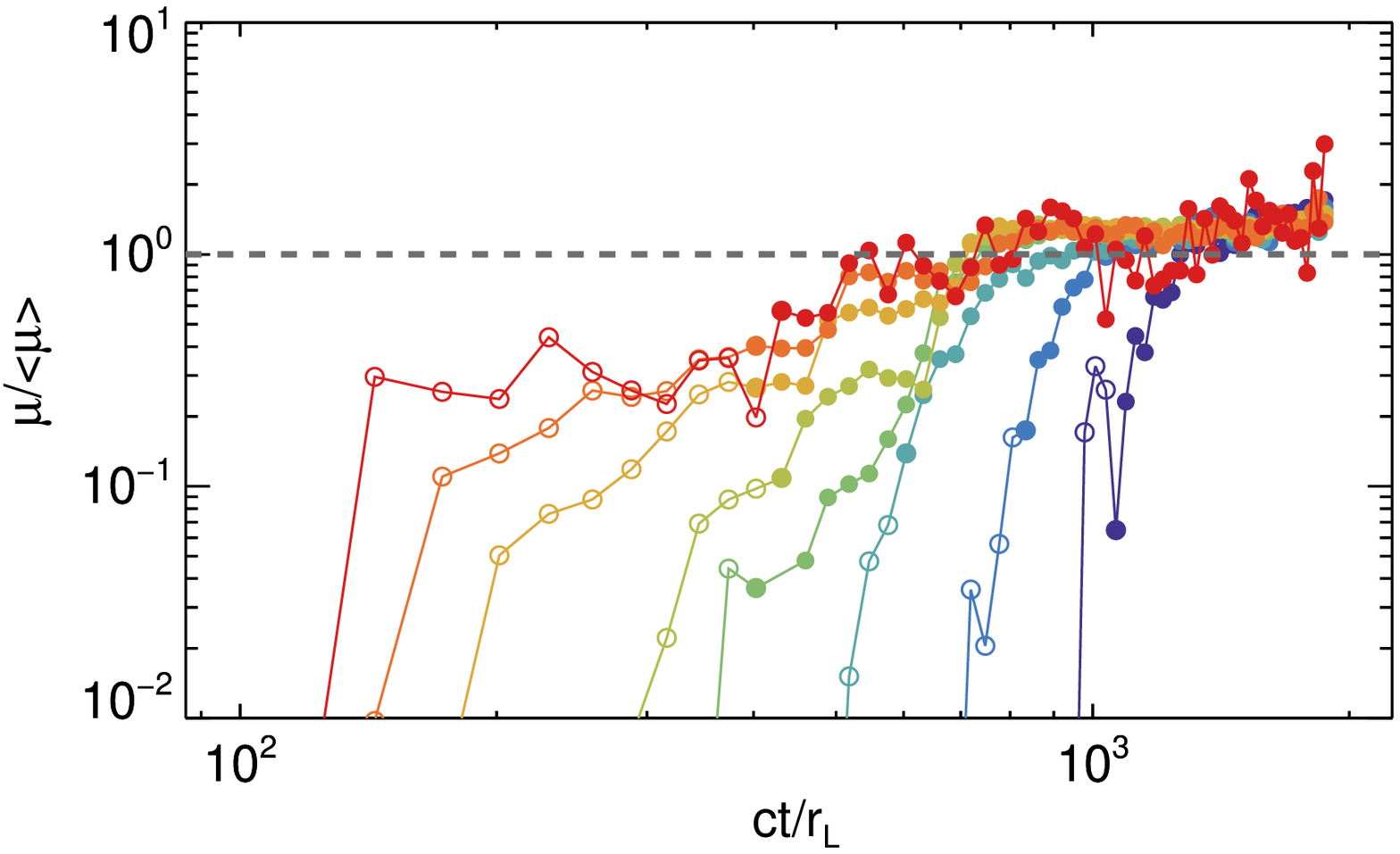}
 \caption{Temporal evolution of various quantities for 1657 positrons that get trapped in the isolated plasmoid $P_1$ shown in \fig{posttime}. Coloured lines with symbols show the median value of a given quantity computed using sub-samples of particles (see text for details).
  From top to bottom we show the temporal evolution of the median value of: the Lorentz factor $\gamma$, the magnetic field strength $B$ at the particle location normalised to its time-averaged value, and the particle magnetic moment $\mu$ normalised to its time-averaged value. In the top panel, the dashed line scales as $\propto \sqrt{t}$, while the dotted one as $\propto t$. In the middle panel, the dashed line scales as $\propto t$. In all the panels, open (filled) symbols denote the times when the median distance (with respect to the plasmoid centre) of particles within a given group is larger (smaller, respectively) than the plasmoid transverse size. In other words, filled circles indicate when the particles reside in the plasmoid.
 }
 \label{fig:nomajor}
\end{figure}
In order to check how representative is the particle trajectory displayed in \fig{multi_prt140}, we computed $\gamma$, $B$, and $\mu$ for 1657 particles that get confined in plasmoid $P_1$ and that possess high energies ($\gamma \geq 4\sigma$) at $t\sim2000 \,\rL/c$ (i.e., when the plasmoid ceases to be isolated). For each particle, we identified the time when its Lorentz factor first exceeds the value of 1.1; this roughly corresponds to the ``injection'' time, when the particle first interacts with the reconnection layer. We then grouped the particles depending on their injection time, using eight equally-spaced logarithmic time bins from  $86\, \rL/c$ up to $1.7\times10^3\, \rL/c$. In \fig{nomajor} we present the temporal evolution of the median values of $\gamma$, $B$, and $\mu$ for each particle group (each group corresponds to a different color). In all panels, open circles are used when the median distance of a group of particles from the centre of $P_1$ exceeds the plasmoid transverse size; otherwise, filled symbols are used (so, filled circles indicate times when more than half of the particles reside in $P_1$). 

The median $\gamma$ (top panel) of the groups of particles that get trapped in $P_1$ at early times (i.e., red to green curves) follows the $\sqrt{t}$ scaling, in analogy to what is found for individual particles (\fig{multi_prt140}) and for the high-energy cutoff $\gc$ of the overall particle energy spectrum (see e.g., \fig{boxsize} and \fig{islspec}). A faster temporal increase of $\gamma$ is found for particles that enter into $P_1$ at later times (i.e., cyan to blue curves), but this is just a consequence of our selection criterion, since our particles have to exceed $\gamma \geq 4\sigma$ at $t\sim2000 \,\rL/c$. During the phase when $P_1$ is isolated ($700\, \rL/c\lesssim t\lesssim 2000\, \rL/c$), the median magnetic field of each particle group shows a linear increase with time (middle panel), while the magnetic moment remains roughly constant (bottom panel).

The results presented in \fig{multi_prt140} and \fig{nomajor} paint the following physical picture for particle energisation in plasmoid-dominated reconnection. During the time in between major island mergers (in what we have called the ``isolated phase''), most of the high-energy particles reside close to island cores and their first adiabatic invariant is conserved. Meanwhile, the magnetic field in island cores increases due to the continuous accretion of field lines as the plasmoid gets bigger, thus leading to an increase of the particle perpendicular momentum. The perpendicular momentum, as well as the particle energy, increases as $\propto \sqrt{t}$.  Mergers act differently for particles residing in the plasmoid outskirts or near the cores. In the early stages of a merger, the relative speed between the two merging plasmoids is large (and so, the anti-reconnection electric field at the merging interface is strong), leading to acceleration of the particles that reside in the plasmoid outskirts (see also \fig{hrmap} and the related discussion). By the time the two island cores merge, their relative speed is much smaller, and the particles residing in the cores do not appreciably change their energy. Yet, mergers can have the effect of moving the high-energy particles residing near island cores towards regions of lower magnetic field, thus allowing for further particle energisation via compression. Thus, mergers provide the ground for multiple energisation/compression cycles.

\section{Summary and Discussion}\label{sec:discussion}
We have studied the long-term evolution of the energy spectrum of particles accelerated by relativistic reconnection in pair plasmas. Using a suite of large-scale 2D PIC simulations we demonstrated that the high-energy cutoff of the particle spectrum does not saturate at $\gamma_{\rm cut}\sim 4\sigma$, as claimed by earlier works. Rather, it steadily grows to significantly larger values as long as the reconnection process stays active (see \fig{spec}). The evolution is fast at early times when $\gamma_{\rm cut}\lesssim 4\sigma$. At later times, the cutoff scales approximately as $\gamma_{\rm cut}\propto \sqrt{t}$, regardless of the flow magnetization and the temperature of the pre-reconnection plasma (see figures in \sect{sec:results} and Appendix~\ref{sec:app1}). The slower evolution of the high-energy spectral cutoff at late times might have been {misinterpreted} as a saturation at $\gamma_{\rm cut}\sim 4\sigma$ in earlier studies \citep{werner_16, kagan_18} that employed smaller computational domains.
 
We demonstrated that the scaling $\gc \propto \sqrt{t}$  holds not only for the energy spectrum of the whole reconnection region, but also for the spectrum extracted from  isolated plasmoids, i.e., those that do not undergo mergers with plasmoids of comparable sizes (see \fig{posttime} and \fig{islspec}). This finding suggests that mergers are not the main drivers of the growth of the high-energy spectral cutoff. By performing  ``tomography'' on an isolated plasmoid, we showed that that hardness of the particle energy spectrum depends on the location within the plasmoid. In particular, the particles dominating the high-energy spectral cutoff reside in a strongly magnetized ring around the plasmoid core, as shown in Figs.~\ref{fig:eBnomajor}, \ref{fig:tomoisl},  and \ref{fig:hrmap}. By following the trajectories of a large number of high-energy particles, we found that the growth of their energy (and consequently, of the spectral cutoff) is driven by the increase in magnetic field at the particle location --- as the plasmoid where they reside compresses while accreting particles and magnetic flux ---  coupled with the conservation of the first adiabatic invariant (see e.g. \fig{eBnomajor}, \fig{multi_prt140}, and \fig{nomajor}). This gradual  energisation mechanism might have been missed by \cite{nalewajko_15}, since the criterion they used for the identification of acceleration sites was biased in favour of fast energisations (see Sect.~3.2.2 therein). 
 
{The formation of a magnetized ring around the core of primary plasmoids is a key ingredient of the proposed particle energisation process. Although the core of primary plasmoids bears memory of the initial conditions in the current sheet, the ring radius is significantly larger than the initial current sheet thickness (see e.g. Table~\ref{tab:param} and Fig.~\ref{fig:eBnomajor}), thus making the initial conditions in the current sheet irrelevant for the proposed particle acceleration mechanism. This is further supported by the fact that the long-term evolution of the high-energy cutoff of the particle distribution is similar for different current sheet thicknesses (see Fig.~\ref{fig:thickness}).}
 
Our analysis also showed that the power-law slope of the particle energy spectrum ($p=-{\rm d}\log N/{\rm d}\log \gamma$) softens over time. For $\sigma=10$, in particular, it asymptotes to $p\sim 2$, corresponding to equal energy content per logarithmic interval in Lorentz factor (see \fig{boxsize}). For $\sigma=50$, the power law index grows with time from $p\sim1.2$ up to $p\sim 1.7$ (see \fig{sig50}). Although not explicitly shown here, it is likely that $p\rightarrow 2$ at even later times. The steepening of the power-law slope observed for $\sigma=50$ allows  the spectral cutoff to extend to higher and higher energies with time (see \eq{gsig}), without violating the fixed energy budget of the system \citep[in contrast to previous claims, e.g.][]{kagan_18}. 

{We demonstrated that plasmoid compression is the main driver of the steady growth of the high-energy spectral cutoff in 2D simulations of relativistic reconnection. Although this reminds of the scenario proposed by \cite{drake_06}, where particle acceleration takes place inside contracting islands, there are some crucial differences that we discuss below. According to \cite{drake_06}, electrons reflect from the two sides of a contracting island, thus increasing their energy (more specifically, the component of momentum parallel to the magnetic field). Particles can therefore accelerate as long as the island aspect ratio changes with time. Instead, we showed that the key ingredient for particle energisation within plasmoids is the continuous accretion of magnetic flux
and plasma. As a related point, \cite{drake_06} invoked the conservation of the second adiabatic invariant to relate the shrinking of the particle path to the particle energisation. The relevant conserved quantity in our case is, however, the first adiabatic invariant, since particles increase their energy mainly in the direction perpendicular to the local magnetic field.} 

Before discussing the astrophysical implications of our findings, we present a few caveats.
We based our analysis of the cutoff evolution on the assumption that the particle energy spectrum can be well described (above a certain Lorentz factor) by a power law with an exponential cutoff (see eq.~(\ref{eq:dndg})). However, we also checked whether a power law followed by a super-exponential cutoff \citep[see e.g.][]{kagan_18} provides a better description of the data.  We found that the simple exponential cutoff is a good model as long as we focus on the long-term evolution of the particle energy spectrum (see inset in \fig{spec}). At early times, there is a degeneracy between models with simple- or super- exponential cutoffs because of the limited extent of the power-law segment and the low cutoff value. We therefore selected the model with the smallest number of free parameters that can describe the data sufficiently well at all times. 

For direct comparison to earlier works \citep{werner_16, kagan_18}, we have studied untriggered/spontaneous reconnection with periodic boundary conditions. However, the region in between two neighbouring primary plasmoids should be physically equivalent to the setup of triggered reconnection with outflow boundaries employed in \citet{sironi_16}. These authors showed that, in fact, the maximum Lorentz factor of accelerated particles in simulations with outflow boundaries steadily increases with the size of the computational domain, which is equivalent  to the steady growth of the cutoff as a function of time found here. It remains to be assessed whether the scaling of the high-energy cutoff seen in  \citet{sironi_16} is consistent with the $\sqrt{t}$ scaling found here. It is possible that the dynamics and structure of secondary plasmoids, which contain the highest energy particles in the simulations of \citealt{sironi_16}, are different from those of the primary plasmoids studied here, thus resulting in a different temporal scaling of the cutoff.

In this work, we have identified plasmoid compression as the main driver of the steady growth of the high-energy spectral cutoff in 2D simulations of reconnection. This slow energisation process is allowed to operate as long as particles are confined in the inner magnetized regions of plasmoids. {However, particle acceleration to very high energies does not require infinitely large magnetic fields, which may be hard to be realised on astrophysical scales. Instead, particles can reach very high energies by experiencing multiple cycles of energization via compression. This is made possible by plasmoid mergers that may kick-off particles from the inner regions to the outer regions of lower magnetic field. Particles may then enter another acceleration/compression cycle, as demonstrated in Fig.~\ref{fig:multi_prt140}.} 

The long-term evolution of the accelerated particle distribution may differ in 3D systems, if other channels for acceleration prevail over the energisation by compression that dominates in our 2D simulations.
In 3D, the compression-driven growth of the high-energy cutoff could also be limited if particles were able to escape the plasmoid (in the $z$-direction) on a timescale much shorter than the plasmoid lifetime. We can estimate the time it would take a particle to escape as it drifts along $z$, assuming that the plasmoid size in the $z$-direction is similar to its transverse size $w_\perp$. Since our high-energy particles have perpendicular momentum larger than the parallel momentum, the grad-B drift will be faster than the curvature drift. Taking the plasmoid size $w_\perp$ as the characteristic scale of the field gradient, the typical escape time for a particle with Larmor radius $r_{\rm g}=\gamma mc^2 / e B$ will be $t_{\rm esc} \approx w^2_{\perp}/c r_{\rm g} = w^2_{\perp} \sqrt{2 \sigma \eB}/c \gamma \rL$, where we used  the relation $B_0/B=\sqrt{\sigma/2 \eB}$. For a particle of $\gamma\sim 50$ located in the highly magnetized ring of the plasmoid $P_1$ with $\eB\sim 20$ and $w_{\perp}\sim 100 \, \rL$ (see Figs.~\ref{fig:eBnomajor} and \ref{fig:hrmap}), we find $t_{\rm esc}\approx 4\times10^3 \, \rL/c$. Our results for the evolution of the high-energy spectral cutoff can be therefore trusted up to the duration of our simulations. {It is also likely that a high-energy particle propagating along the $z$-direction due to the drift, will encounter the edge of a plasmoid (i.e., regions of lower magnetic field) and will start sampling the two sides of the layer on a Speiser orbit \citep{giannios_10}; this will lead to additional (Fermi-like) acceleration with energy increase linear in time.}

 Our findings have important implications for astrophysical non-thermal sources.
Electrons radiating in a variety of astrophysical outflows must be accelerated to ultra-relativistic energies in order to produce the observed non-thermal emission ($\gamma \sim 10^3 - 10^6$ in blazars, see e.g. \citealt{celotti_08, petro_dim_pado_15}; up to $\gamma\sim 10^9$ in the recently discovered gamma-ray flares from the Crab nebula, \citealt{buehler_12}). Saturation of the high-energy spectral cutoff at $\gc \sim 4\sigma$ (as argued in earlier studies) would therefore pose a major problem for reconnection-powered models of particle acceleration, as it would require a relativistically hot ($\Theta_{\rm e}\gg 10^2$) or/and a highly magnetized ($\sigma \gg 10^2$) pre-reconnected plasma in virtually all astrophysical sources; this is extremely unlikely. We have shown that these strict requirements are not necessary for the reconnection scenario to work, since particles can be accelerated by reconnection to Lorentz factors well beyond  $4\sigma$.

We have argued that the asymptotic slope of the accelerated particle spectrum in plasmas with $\sigma \gtrsim 10$ is $p \sim 2$. At first, this might seem in tension with observations of blazars having hard $\gamma$-ray spectra, since these require the accelerated electrons to have also hard  spectra with $p<2$ \citep[e.g.,][]{aharonian_07, lefa_11, neronov_12}. However, we expect $p \sim 2$ only for an asymptotically-large system.
In a realistic system of finite extent, the duration of a reconnection event depends on the available magnetic flux and the expansion timescale of the current sheet (so, it is set by the global dynamics of the system). Thus, it is likely that the finite duration of the reconnection episode is not sufficient to establish the asymptotic $p\sim 2$ slope. In this regard, reconnection in plasmas with $\sigma > 10$ is still a viable (and one of the few) physical mechanism for the production of hard particle spectra.

\section{Conclusion}\label{sec:concl}
Based on this work, we argue that a unified picture of particle acceleration in relativistic magnetic reconnection seems to emerge. Particles are injected into the acceleration process when they first encounter the magnetic X-points of the current sheet. Then, they are advected into plasmoids, where they can be further energised, albeit at a slower rate, by plasmoid compression. Mergers occurring during the lifetime of a plasmoid can also contribute to the acceleration of selected particles, predominantly of those residing at the outskirts of the merging plasmoids. Nevertheless, the steady growth of the high-energy spectral cutoff is controlled by plasmoid compression and therefore holds even in the absence of major mergers.
 
\section*{Acknowledgments}
The authors would like to thank the anonymous referee for a constructive report and Drs. L.~Comisso, D.~Giannios, D.~Kagan, and J.~Zrake for useful comments.
The authors also thank Prof. A. Spitkovsky for many fruitful discussions. MP acknowledges support from the Lyman Jr.~Spitzer Postdoctoral Fellowship. LS acknowledges support from DoE DE-SC0016542, NASA Fermi NNX-16AR75G, NASA ATP NNX-17AG21G, NSF ACI-1657507, and NSF AST-1716567. The simulations were performed on Habanero at Columbia, on NASA (Pleiades) and NERSC (Edison) resources.

\bibliographystyle{mnras} 
\bibliography{blob}

\appendix

\section{Dependence on Numerical Parameters}\label{sec:app1}
\begin{figure}
\centering
 \resizebox{\hsize}{!}{\includegraphics[trim=0 20 0 0, clip=true]{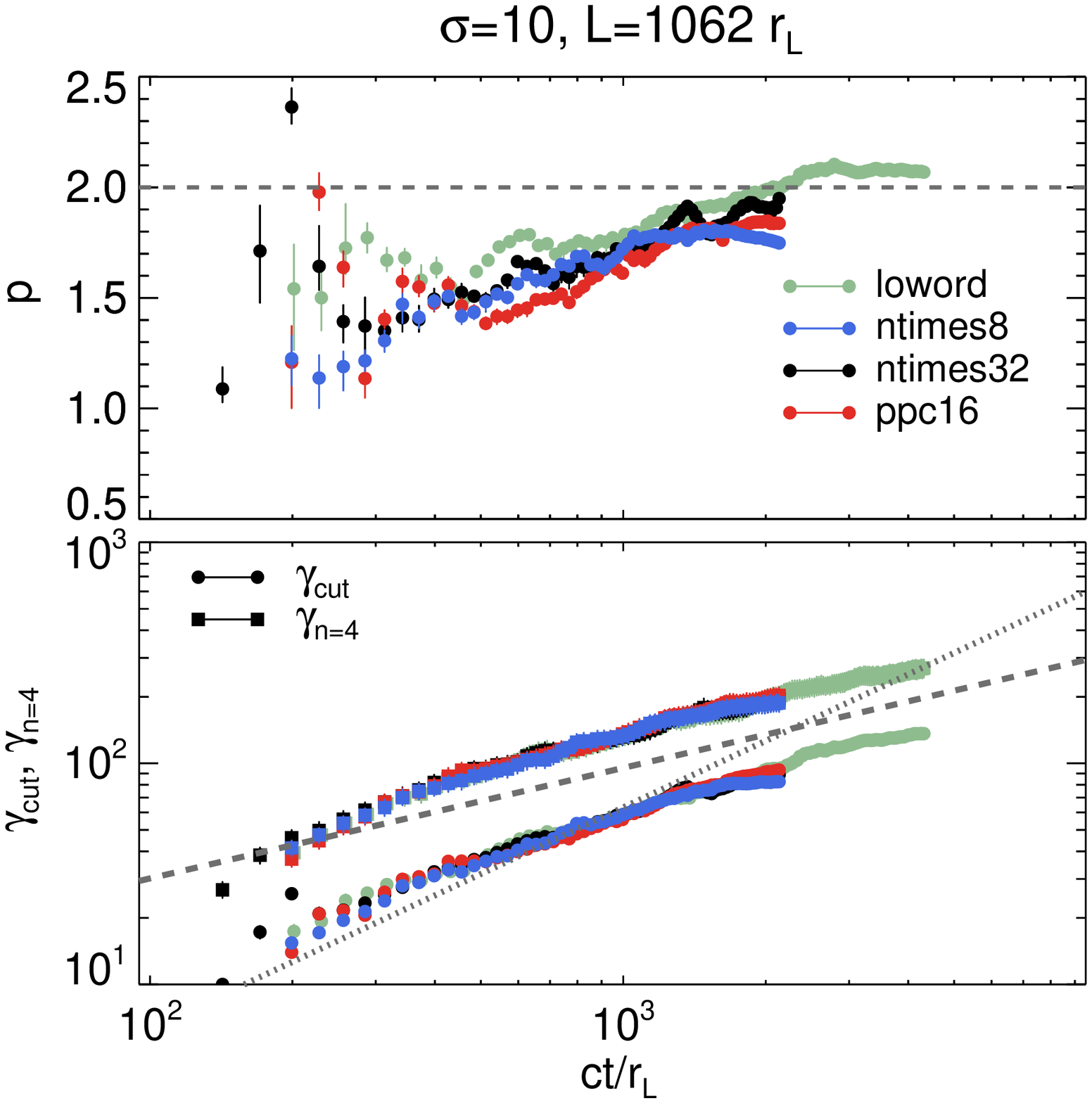}}
 \caption{Same as in \fig{boxsize}, but for different numerical parameters of the simulations (see also Table~\ref{tab:param}). 
 }
 \label{fig:num}
\end{figure}

\begin{figure}
\centering
 \resizebox{\hsize}{!}{\includegraphics[trim=0 20 0 0, clip=true]{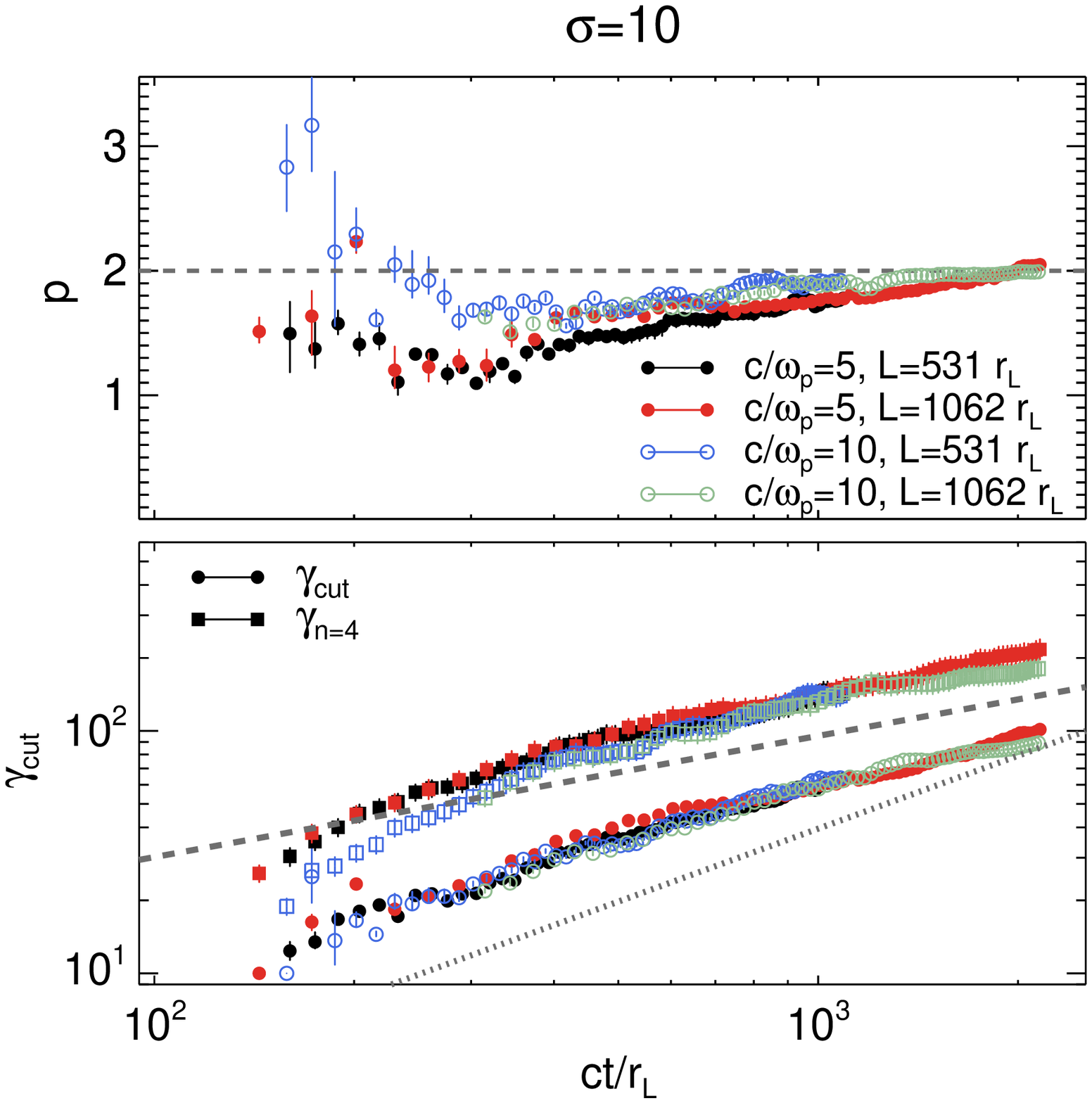}}
 \caption{Same as in \fig{boxsize}, but for different numerical resolutions: $\comp=5$ (filled symbols) and $\comp=10$ (open symbols).
 }
 \label{fig:comp}
\end{figure}

In order to test the robustness of our results about the long-term spectral evolution, we performed a limited number of simulations (\tab{param}), where we kept the same physical parameters (e.g., $\sigma$ and $\Theta_{\rm e}$), but varied several numerical parameters. These are:
the integration scheme for Maxwell's equations (``loword'', based on a second-order spatial stencil \citep{yee_66}, rather than the fourth-order stencil \citep{greenwood_04} employed in the main body of this paper), the number of particles per cell (``ppc16''), the number of filter passes on the electric currents (``ntimes8'' and ``ntimes32''), and the numerical resolution ($\comp$). The temporal evolution of the slope $p$ and the high-energy cutoff $\gc$ are summarised in \fig{num} and \fig{comp}. Overall, we find that none of the numerical parameters tested here affects the results presented in \sect{sec:results}.

\end{document}